\documentclass[twocolumn,aps,showpacs]{revtex4}
\usepackage{amsfonts}
\usepackage{amsmath}
\usepackage{graphicx}
\usepackage{bm}
\usepackage{amssymb}
\usepackage{dcolumn}

\newcommand{\ltsim}{\protect\raisebox{-0.5ex}{$\:\stackrel{\textstyle <}{\sim}\:$}}
\newcommand{\gtsim}{\protect\raisebox{-0.5ex}{$\:\stackrel{\textstyle >}{\sim}\:$}}

\begin{document}

\title{Inter-impurity and impurity-host magnetic correlations in semiconductors\\
with low-density transition-metal impurities}

\author{Yoshihiro Tomoda$^{1}$, Nejat Bulut$^{1,2}$, and Sadamichi Maekawa$^{1,2}$}

\affiliation{$^1$ Institute for Materials Research, Tohoku
University, Sendai 980-8577, Japan\\ $^{2}$CREST, Japan Science
and Technology Agency (JST), Kawaguchi, Saitama 332-0012, Japan}

\date{May 31, 2008}

\begin{abstract}
Experiments on (Ga,Mn)As in the low-doping insulating phase have
shown evidence for the presence of an impurity band at 110 meV
above the valence band. The motivation of this paper is to
investigate the role of the impurity band in determining the
magnetic correlations in the low-doping regime of the dilute
magnetic semiconductors. For this purpose, we present results on
the Haldane-Anderson model of transition-metal impurities in a
semiconductor host, which were obtained by using the Hirsch-Fye
Quantum Monte Carlo (QMC) algorithm. In particular, we present
results on the impurity-impurity and impurity-host magnetic
correlations in two and three-dimensional semiconductors with
quadratic band dispersions. In addition, we use the tight-binding
approximation with experimentally-determined parameters to obtain
the host band structure and the impurity-host hybridization for Mn
impurities in GaAs. When the chemical potential is located between
the top of the valence band and the impurity bound state (IBS),
the impurities exhibit ferromagnetic (FM) correlations with the
longest range. We show that these FM correlations are generated by
the antiferromagnetic coupling of the host electronic spins to the
impurity magnetic moment. Finally, we obtain an IBS energy of 100
meV, which is consistent with the experimental value of 110 meV,
by combining the QMC technique with the tight-binding approach for
a Mn impurity in GaAs.
\end{abstract}


\maketitle

\section{Introduction}

The discovery of dilute magnetic semiconductor (DMS) materials is
important because of possible spintronics device applications
\cite{Maekawa,Ohno,Zutic,Jungwirth}. The electronic state of the
alloy (Ga,Mn)As, where Mn substitutes for Ga, has been
investigated by various experimental methods including transport
measurements, optical and photoemission spectroscopy, and
scanning-tunnelling microscopy
\cite{Chapman,Blakemore,Okabayashi,Burch,Yakunin,Kitchen}. In the
low-doping regime ($\ll 1\%$), this prototypical DMS material is
insulating with clear experimental evidence for the presence of an
Mn-induced impurity band located 110 meV above the valence band.
For more than 2\% Mn doping, the dc transport measurements
indicate that the impurity band emerges with the valence band
leading to a metallic state. The redshift of the mid-infrared peak
observed in optical absorption in metallic samples has been
attributed to inter-valence band transition instead of being due
to an impurity band \cite{Burch,Jungwirth2}. Hence, the analysis
of the experimental data suggest that (Ga,Mn)As exhibits two
different regimes at low and high impurity concentrations
separated by an insulator-metal transition \cite{Jungwirth2}.

In this paper, our purpose is to develop a microscopic
understanding of the low-doping insulating phase of the DMS
material (Ga,Mn)As and to shed light on the role of the impurity
band in producing the magnetic correlations. For this purpose, we
present numerical results on the single- and two-impurity
Haldane-Anderson model of transition-metal impurities in
semiconductors \cite{Haldane} using the Hirsch-Fye Quantum Monte
Carlo (QMC) method \cite{Hirsch}. We study the influence of the
electronic state of the semiconductor host on the impurity bound
state (IBS) and, in turn, on the magnetic correlations which
develop between the impurities and between the impurity and the
host electrons. We find that, when the chemical potential $\mu$ is
located between the valence band and the IBS, the ferromagnetic
(FM) correlations between the impurities exhibit the longest
range. We also show that the inter-impurity FM correlations are
generated by the antiferromagnetic (AFM) correlations between the
impurities and the host electrons. In addition, we use the
tight-binding approximation to determine the host band structure
and the impurity-host hybridization for the three $t_{2g}$
orbitals of Mn in GaAs. Using tight-binding parameters consistent
with photoemission measurements, we obtain an IBS energy which is
in agreement with the experimental value of 110 meV. Because of
these results, we think that a microscopic understanding of the
low-doping insulating phase of (Ga,Mn)As is possible within the
Haldane-Anderson model.

The nature of the magnetic correlations in the Haldane-Anderson
model of transition-metal impurities in semiconductors was studied
using the Hartree-Fock (HF) approximation
\cite{Ichimura,Takahashi}. It was shown that, when the chemical
potential is located between the top of the valence band and the
IBS located at $\omega_{IBS}$, long-range FM correlations develop
between the impurities mediated by the AFM coupling of the valence
electrons to the impurity magnetic moments. On the other hand,
when the IBS becomes occupied, the spin polarization of the host
split-off state cancels the polarization of the valence band,
reducing the range of the FM correlations. The QMC calculations
for the Haldane-Anderson model support the HF picture for the role
of the IBS in producing the FM correlations \cite{Bulut}. The QMC
results on the range of the FM correlations are in agreement with
the HF predictions, however, the HF approximation underestimated
the value of $\omega_{IBS}$. These results on the magnetic
correlations for $0 \lesssim \mu \lesssim \omega_{IBS}$ are
different than in the metallic case, $\mu < 0$, where the
inter-impurity magnetic correlations exhibit
Ruderman-Kittel-Kasuya-Yosida (RKKY) oscillations with period
determined by the Fermi wavevector $k_F$. We note that the
Haldane-Anderson model was also studied within HF by Krstaji\'c
{\it et al.} \cite{Krstajic} for DMS materials and by Yamauchi
{\it et al.} {\cite{Yamauchi} for hemoprotein. The role of IBS for
DMS materials was also discussed by Inoue {\it et al.}
\cite{Inoue} within HF.

In this paper, we are particularly interested in how the magnetic
properties are influenced by the host electronic state, which we
model using two different approaches. First, we consider the
simple case where the magnetic impurity has one orbital only and
the host band structure consists of quadratic valence and
conduction bands. Here, in addition, we treat the impurity-host
hybridization as a freely adjustable parameter. This
single-orbital model is described in Section 2.A, and the QMC
results for this case are presented in Sections 3 and 4 for two
and three-dimensional semiconductors, respectively. We study the
inter-impurity and impurity-host magnetic correlations, the
induced electron-density around the impurity, and show how these
quantities depend on the parameters of the Haldane-Anderson model
and the dimensionality of the semiconductor. These results show
that the IBS and the magnetic correlations depend sensitively on
the model parameters.

In order to develop a more realistic model for (Ga,Mn)As, we next
use the tight-binding approximation to obtain the band structure
of the bare host GaAs and the hybridization matrix elements with
the three $t_{2g}$ orbitals of Mn substituted in place of Ga. In
this approach, the tight-binding parameters required for
calculating the hybridization are taken from an analysis of the
photoemission data on Mn $2p$ core level. This model is introduced
in Section 2.B and the QMC results obtained for this case are
presented in Section 5. These QMC calculations keep all three of
the Mn $t_{2g}$ orbitals, and hence the multi-orbital effects are
included except for those due to Coulomb repulsion between
different Mn orbitals. Within this approach, we obtain an
$\omega_{IBS}$ which is close to the experimental value of 110
meV. We also show that the FM correlations between the impurities
weaken as the IBS becomes occupied. These results emphasize the
importance of the IBS in the low-density limit of (Ga,Mn)As.

Finally, we note that the approaches taken in this paper can be
extended to the case of finite concentration of Mn impurities in
order to study the insulator-metal transition and the metallic
state of (Ga,Mn)As observed at higher Mn concentrations.

\section{Impurity Model}

The general model for describing transition-metal impurities in a
semiconductor host is given by Haldane-Anderson Hamiltonian
\cite{Haldane},
\begin{eqnarray}
H &=& \sum_{{\bf k},\alpha,\sigma} (\varepsilon_{{\bf
k}\alpha}-\mu) c^{\dagger}_{{\bf k}\alpha\sigma} c_{{\bf
k}\alpha\sigma} + \sum_{i,\xi,\sigma} (E_{d\xi}-\mu)
d^{\dagger}_{i\xi\sigma}
d_{i\xi\sigma} \nonumber \\
&+& \sum_{{\bf k},\alpha,i,\xi,\sigma} (V_{{\bf k}\alpha,i\xi}
c^{\dagger}_{{\bf k}\alpha\sigma} d_{i\xi\sigma} +
{\rm H.c.}) \nonumber \\
&+& U\sum_{i,\xi} d^{\dagger}_{i\xi\uparrow}d_{i\xi\uparrow}
d^{\dagger}_{i\xi\downarrow}d_{i\xi\downarrow},
\end{eqnarray}
where $c^{\dagger}_{{\bf k}\alpha\sigma}$ ($c_{{\bf
k}\alpha\sigma}$) creates (annihilates) a host electron with
wavevector ${\bf k}$ and spin $\sigma$ in the valence or
conduction bands denoted by $\alpha$, and
$d^{\dagger}_{i\xi\sigma}$ ($d_{i\xi\sigma}$) is the creation
(annihilation) operator for a localized electron at impurity
orbital $\xi$ located at site $i$. The first term in Eq. (1)
represents the kinetic energy of the host electrons, and the
second term is the bare energy of the localized impurity orbitals,
while the third term is due to the impurity-host hybridization.
The last term represents the onsite Coulomb repulsion at the
impurity orbitals. We note that here we are neglecting the Coulomb
repulsion among the different impurity orbitals. The effects of
the Hund couplings will be considered in a separate paper. As
usual in Eq. (1), $U$ is the onsite Coulomb repulsion, $\mu$ the
chemical potential and $E_{d\xi}$ is the bare energy of the
impurity orbital $\xi$. In addition, the hybridization matrix
element is
\begin{equation}
V_{{\bf k}\alpha,j\xi} = V_{{\bf k}\alpha,\xi}  e^{i {\bf k}\cdot
{\bf R}_j},
\end{equation}
where ${\bf R}_j$ is the coordinate of the impurity site $j$ and
$V_{{\bf k}\alpha,\xi}$ is the value when the impurity is located
at the origin. We use the Hirsch-Fye QMC technique to study the
Haldane-Anderson impurity Hamiltonian, Eq. (1), for the the
single- and two-impurity cases \cite{Hirsch}.

In this paper, we perform the QMC calculations using two different
approaches for incorporating the host band structure and the
impurity-host hybridization. In Sections 3 and 4, we consider a
simple model where the impurity site contains a single impurity
orbital and the semiconductor bands have quadratic dispersion,
while in Section 5 we use the tight-binding approximation for
treating the three $t_{2g}$ orbitals of Mn impurities in GaAs. In
the remainder of this section, we describe these two different
approaches for modelling the transition-metal impurities in
semiconductors.

\subsection{Single-orbital case}

In Sections 3 and 4, we consider the simple case where the
impurity site contains a single orbital in two and
three-dimensional semiconductor hosts, respectively. Hence, in
this case Eq. (1) reduces to Eq. (1) of Ref.~\cite{Bulut}.
Furthermore, in these sections, we assume that the host band
structure consists of one valence ($\alpha=v$) and one conduction
($\alpha=c$) bands with quadratic dispersions given by
\begin{eqnarray}
\varepsilon_{{\bf k},v} &&= - D \left( k / k_0 \right)^2 \\
\varepsilon_{{\bf k},c} &&= D \left( k / k_0 \right)^2 + \Delta_G.
\end{eqnarray}
Here, $D$ is the bandwidth, $k_0$ the maximum wavevector and
$\Delta_G$ the semiconductor gap. The energy scale is determined
by setting $D=12.0$. Figure 1 shows a sketch of the host band
structure. For the Coulomb repulsion we use $U=4.0$, and set the
bare value of the impurity orbital energy to $E_d=\mu-U/2$, so
that the impurity sites develop large magnetic moments. We report
results for semiconductor gap $\Delta_G=2.0$, and inverse
temperature $\beta\equiv 1/T$ from 4 to 32. We also use a constant
$V$ for $V_{{\bf k}\alpha,\xi}$, and treat $V$ as a free
parameter.

\begin{figure}[t]
\includegraphics[width=4.5cm,bbllx=209,bblly=255,bburx=440,bbury=540,clip]{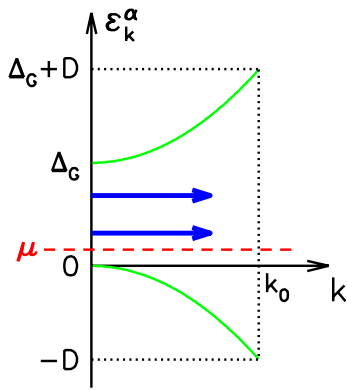}
\caption{(Color online) Schematic drawing of the semiconductor
host bands $\varepsilon_k^{\alpha}$ (solid curves) and the
impurity bound states (thick arrows) obtained with HF in the
semiconductor gap. The dashed line denotes the chemical potential
$\mu$. }\label{fig1}
\end{figure}

For the single orbital cases of Sections 3 and 4, we show QMC
results on the equal-time impurity-impurity magnetic correlation
function $\langle M_1^z M_2^z\rangle$ for the two-impurity
Haldane-Anderson model. Here, the impurity magnetization operator
is given by
\begin{equation}
M_i^z = d^{\dagger}_{i\uparrow}d_{i\uparrow} -
d^{\dagger}_{i\downarrow}d_{i\downarrow},
\end{equation}
and the fermion creation and annihilation operators act at a
single orbital at the impurity site. In addition, we present
results on the impurity-host correlation function $\langle M^z
m^z(r)\rangle$ for the single-orbital and single-impurity
Haldane-Anderson model. Here, the host magnetization at a distance
$r$ away from the impurity site is given by
\begin{equation}
m^z(r) = \sum_{\alpha=v,c} (c^{\dagger}_{\alpha\uparrow}(r)
c_{\alpha\uparrow}(r)-c^{\dagger}_{\alpha\downarrow}(r)
c_{\alpha\downarrow}(r)).
\end{equation}
For the metallic case, the correlation functions $\langle M_1^z
M_2^z\rangle$ and $\langle M^z m^z(r)\rangle$ were previously
studied by using QMC \cite{Hirsch,Fye,Fye2,Gubernatis}. In the
single impurity case, we also present QMC data on the square of
the impurity moment, $\langle (M^z)^2\rangle$, and the impurity
susceptibility $\chi$ defined by
\begin{equation}
\chi = \int_0^{\beta} d\tau \langle M^z(\tau) M^z(0) \rangle.
\end{equation}
The effects of IBS are clearly visible in these quantities. We
also present QMC data on the charge distribution of the host
material around the impurity. These QMC results are obtained using
Matsubara time step $\Delta\tau=0.25$. At $\beta=16$, $\langle
M_1^z M_2^z\rangle$ varies by a few percent as $\Delta\tau$
decreases from 0.25 to 0.125.

\subsection{Tight-binding model for Mn in GaAs}

In order to construct a more realistic model of (Ga,Mn)As in the
dilute limit, we use the tight-binding approach to calculate
$\varepsilon_{{\bf k}\alpha}$ and $V_{{\bf k}\alpha,i\xi}$ of
Eq.~(1) for the Mn $t_{2g}$ orbitals. In Section 5, we will
present QMC data obtained using these results on
$\varepsilon_{{\bf k}\alpha}$ and $V_{{\bf k}\alpha,i\xi}$.

In this approach, the tight-binding band structure
$\varepsilon_{{\bf k}\alpha}$ of GaAs host is obtained by keeping
one $s$ orbital and three $p$ orbitals at each Ga and As site of
GaAs with the zincblende crystal structure. We consider only the
nearest-neighbor hoppings between the Ga and As sites. Figure 2
shows the resulting band energies $\varepsilon_{{\bf k}\alpha}$
versus ${\bf k}$ along various directions in the
face-centered-cubic (FCC) Brillouin zone for pure GaAs. The
Slater-Koster parameters \cite{Slater} which have been used for
obtaining these results were taken from Ref.~\cite{Chadi}. In Fig.
2, the top of the valence band is located at the $\Gamma$ point,
where the energy gap is 1.6 eV, which is consistent with the
experimental value. This simple approximation, with
nearest-neighbor hopping among the eight $sp^3$ orbitals,
reconstructs the four valence bands reasonably. They are important
for the impurity-host magnetic coupling in (Ga,Mn)As, because the
chemical potential is located near the top of the valence band.

\begin{figure}[t]
\centering
\includegraphics[height=8cm,bbllx=55,bblly=55,bburx=370,bbury=340,clip]{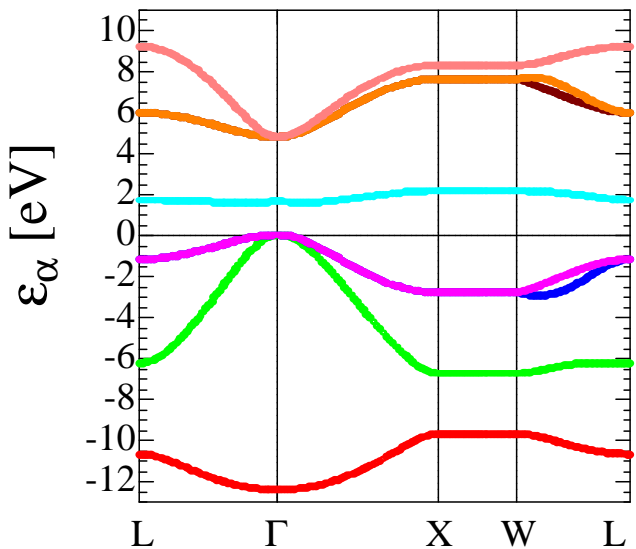}
\caption{ (Color online) Band structure of GaAs within the
tight-binding approximation. } \label{fig2}
\end{figure}

When a Mn impurity is substituted in place of Ga in GaAs, the five
$3d$ orbitals of the Mn ion are split by the tetrahedral crystal
field into the three-fold degenerate $t_{2g}$ orbitals and the
two-fold degenerate $e_{g}$ orbitals. Since the $e_g$ orbitals
have bare energies which are much lower than the $t_{2g}$
orbitals, we neglect the $e_g$ orbitals and keep only the three
$t_{2g}$ orbitals of Mn in Eq.~(1). In order to calculate $V_{{\bf
k}\alpha,i\xi}$ for the Mn $t_{2g}$ orbitals, we assume that the
$sp^3$ orbitals of Mn are the same as those of Ga, and this way
the periodic boundary conditions are also satisfied. In other
words, at the Mn impurity site, we only add the three Mn $t_{2g}$
orbitals to the GaAs host. This is a simple model for a
transition-metal impurity substituted into GaAs. However, this
approach allows us to take into account the effects of the host
band structure beyond the quadratic dispersion. Furthermore, in
Sections 3 and 4, we treat the hybridization matrix element $V$ as
a free parameter. However, here, we perform the calculations using
realistic parameters for the hybridization of the $3d$ orbitals
with the host semiconductor bands. We also note that in
Ref.~\cite{BoGu-ZnO} the tight-binding and QMC techniques have
been combined to study how the crystal structure of ZnO host
affects the magnetic properties when Mn impurities are
substituted. However, in these calculations only one of the Mn
$3d$ orbitals is taken into account and the multi-orbital effects
are not included.

Within the tight-binding approximation, the hybridization matrix
element $V_{{\bf k}\alpha,\xi}$, Eq.~(2), is obtained from
\begin{equation}
V_{{\bf k}\alpha,\xi} = \langle \psi_{{\bf k}\alpha} | H_0 |
\varphi_{\xi} \rangle,
\end{equation}
where $|\psi_{{\bf k}\alpha} \rangle$ is the Bloch eigenstate of
the host electrons, $|\varphi_{\xi}\rangle$  the localized orbital
eigenstate of an impurity located at the origin, and $H_0$ is the
Hamiltonian of the system with the Coulomb repulsion turned off.
In order to evaluate $V_{{\bf k}\alpha,\xi}$, it is necessary to
determine the values of the Slater-Koster parameters between the
$sp^3$ orbitals of GaAs and the $d$ orbitals of Mn, which are
denoted by the notation $(sp\sigma)$, $(pd\sigma)$ and $(pd\pi)$
\cite{Slater}. In the following, we determine the value of
$(pd\pi)$ by the general equation $(pd\pi)=(pd\sigma)/(-2.16)$
\cite{Harrison}. We also set $(sd\sigma)=0$ for the following
reason: When the chemical potential is near the top of the valence
band, the hybridization matrix elements near the $\Gamma$ point
become important for the magnetic couplings. However, the $s$
orbitals do not contribute to the three degenerate valence bands
at the top of the valence band \cite{Tomoda}. This is why we
ignore the contribution of the $s$ orbitals to hybridization and
set $(sd\sigma)=0$. The experimental estimate for the remaining
Slater-Koster parameter is $(pd\sigma)=-1.1$ eV, which is obtained
from the photoemission data on the Mn $2p$ core level with a
configuration-interaction analysis based on a cluster model
\cite{Okabayashi98}. In Section 5, we will present QMC data
obtained using $(pd\sigma)=-1.14$ eV yielding an
$\omega_{IBS}=100$ meV, which is in reasonable agreement with the
experimental value of 110 meV for (Ga,Mn)As in the low-density
limit.

Figure 3 displays tight-binding results on $V_{{\bf k}\alpha,\xi}$
versus ${\bf k}$ for the $\xi=xy$ orbital, and (a) the valence and
(b) conduction bands along various cuts in the FCC Brillouin zone.
The values of $V_{{\bf k}\alpha,\xi}$ at the top of the valence
and at the bottom of the conduction band are particularly
important for the IBS. We note that $V_{{\bf k}\alpha,\xi=xy}$
takes large values near the $\Gamma$ point for the top valence
bands, while it is weaker for the conduction bands. In fact, the
hybridization of the $xy$ orbital with the lowest-lying conduction
band vanishes at the $\Gamma$ point. We also notice that $V_{{\bf
k}\alpha,\xi}$ can be discontinuous around the high-symmetry
points. In Section 5, we will present QMC data obtained using
these values for $\varepsilon_{{\bf k}\alpha}$ and $V_{{\bf
k}\alpha,\xi}$. In addition, for the bare energies of the impurity
orbitals, we will use $E_{d\xi}=\mu-U/2$ with $U=4.0$, so that
large magnetic moments develop at the impurity sites. The QMC
results do not depend sensitively on small changes in the values
of $E_{d\xi}$ and $U$.

\begin{figure}[t]
(a)\includegraphics[height=5cm,bbllx=55,bblly=75,bburx=280,bbury=275,clip]{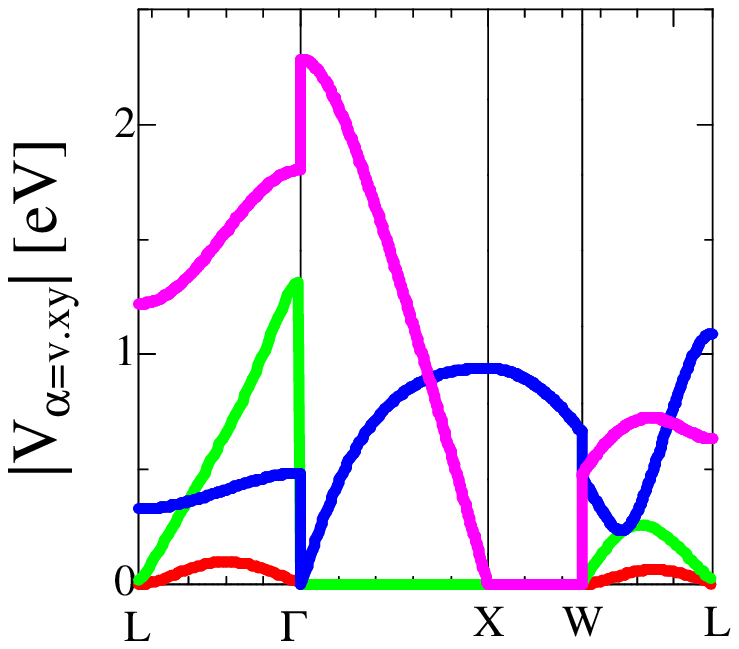}
(b)\includegraphics[height=5cm,bbllx=55,bblly=75,bburx=280,bbury=275,clip]{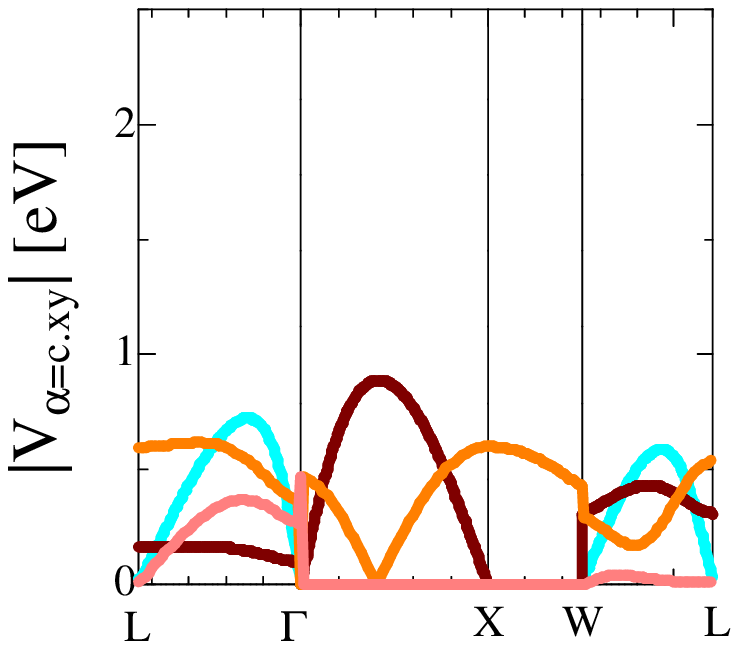}
\caption{ (Color online) Hybridization matrix element $V_{{\bf
k}\alpha,xy}$ of the $\xi=xy$ orbital of a Mn impurity with (a)
the valence and (b) the conduction bands of GaAs versus ${\bf k}$
along various directions in the FCC Brillouin zone obtained using
the tight-binding approximation. } \label{fig3}
\end{figure}

After obtaining $\varepsilon_{{\bf k}\alpha}$ and $V_{{\bf
k}\alpha,i\xi}$ within the tight-binding model described above, we
perform the QMC calculations keeping all three Mn $t_{2g}$
orbitals. Hence, the QMC results which will be presented in
Section 5 include the multi-orbital effects except for the Hund
coupling. However, we will present these QMC results only for the
$\xi=xy$ orbital in order to make comparisons with the the
single-orbital cases of Sections 3 and 4. In particular, defining
the magnetization operator for $\xi=xy$ as
\begin{equation}
M_i^z = d^{\dagger}_{i\xi\uparrow}d_{i\xi\uparrow} -
d^{\dagger}_{i\xi\downarrow}d_{i\xi\downarrow},
\end{equation}
we will present data on $\langle (M^z)^2\rangle$ and the magnetic
susceptibility $\chi$, Eq.~(7), for the $xy$ orbital within the
single-impurity Haldane-Anderson model. For the two-impurity case,
we will present data on the inter-impurity magnetic correlation
function $\langle M_1^z M_2^z \rangle$ between the $xy$ orbitals
located at the impurity sites ${\bf R}_1$ and ${\bf R}_2$. In
addition, we will discuss the magnetic correlation function
between the magnetic moment of the $xy$ orbital and the host
electrons $\langle M^z m^z({\bf r}) \rangle$, where the host
magnetization operator at lattice site ${\bf r}$ is
\begin{equation}
m^z({\bf r}) = \sum_{\alpha} ( c^{\dagger}_{{\bf r}\alpha\uparrow}
c_{{\bf r}\alpha\uparrow} - c^{\dagger}_{{\bf r}\alpha\downarrow}
c_{{\bf r}\alpha\downarrow} )
\end{equation}
with $\alpha$ summing over the eight semiconductor bands, for the
single impurity case. In Section 5, we will see that the
experimental value of $\omega_{IBS}$ is reproduced reasonably by
combining the tight-binding approach with the QMC calculations. In
addition, we will see that the magnetic correlations weaken as the
IBS becomes occupied, in agreement with the results of Sections 3
and 4.

In the following sections, we will find that the quantitative
results depend on the dimensionality and the band structure of the
host material. In Sections 3 and 4, we show results for the two
and three dimensional host materials with simple quadratic band
dispersions. Then, in Section 5, we discuss the case for a GaAs
host using the tight binding approximation.

\section{QMC results  for a 2D semiconductor host}

In this section, we show results for the 2D case. Here, the
density of states of the pure host, $\rho_0$, is a constant with a
sharp cutoff at the semiconductor gap edge, which leads to
stronger impurity-host coupling compared to the 3D case. Here, we
present results for hybridization $\Delta\equiv \pi \rho_0 V^2$
varying from 1 to 4.

\subsection{Magnetic correlations between the impurities}

\begin{figure}[t]
\centering
\includegraphics[width=7cm,bbllx=96,bblly=212,bburx=525,bbury=494,clip]{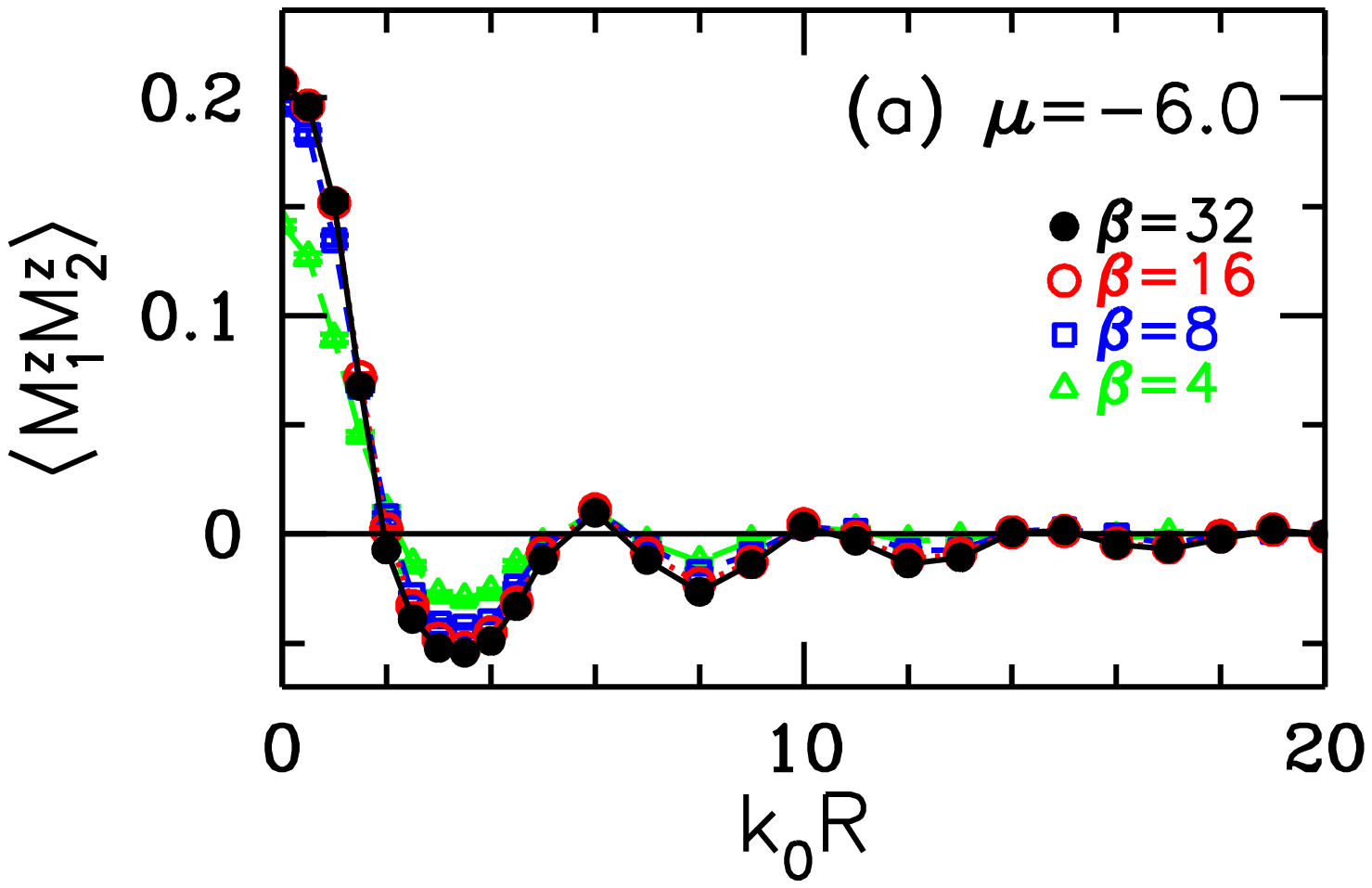}
\includegraphics[width=7cm,bbllx=96,bblly=212,bburx=525,bbury=494,clip]{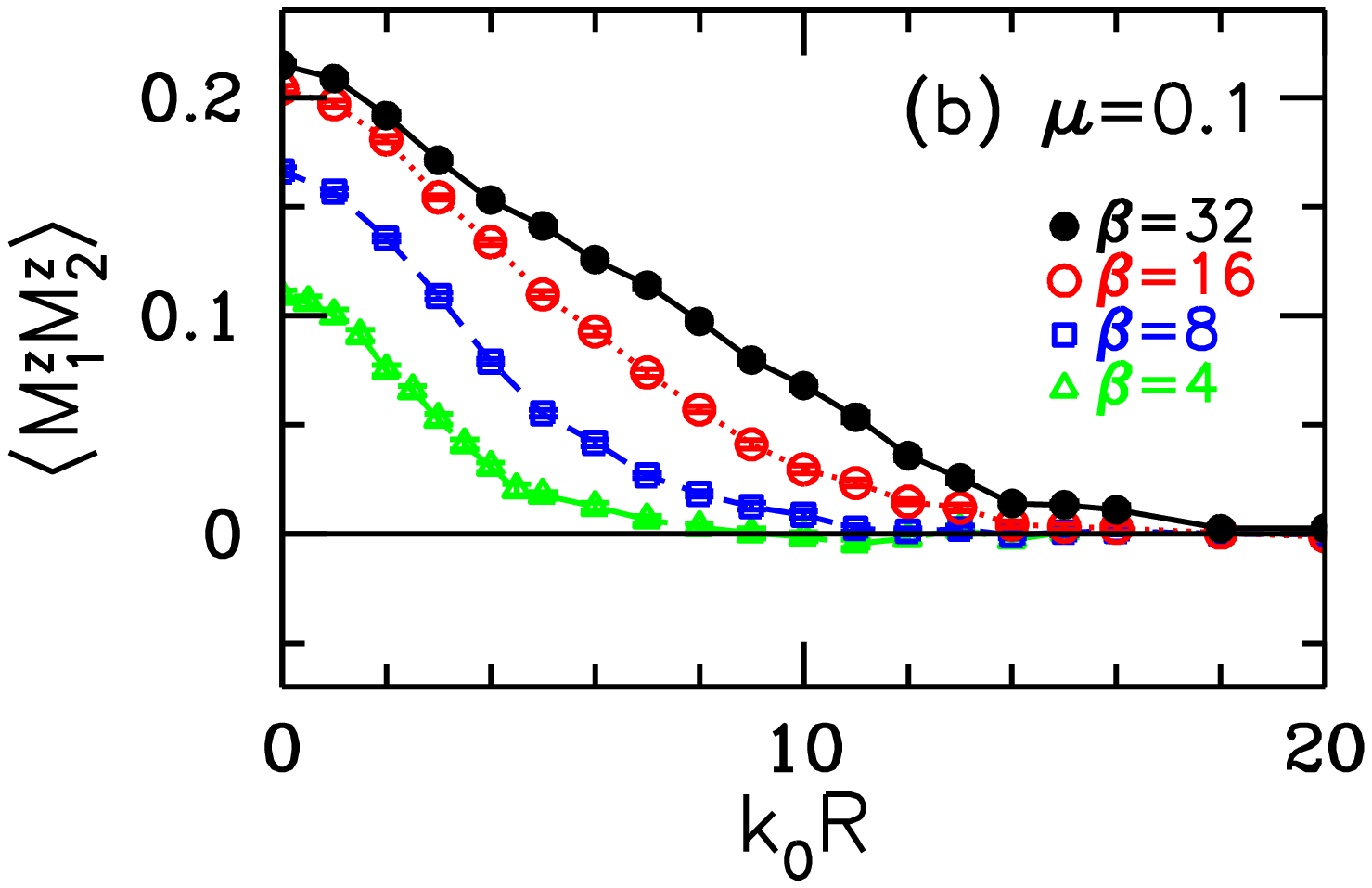}
\caption{ (Color online) Inter-impurity magnetic correlation
function $\langle M_1^z M_2^z\rangle$ vs $k_0R$ at various $\beta$
for (a) $\mu=-6.0$ and (b) $\mu=0.1$ for the two-impurity
Haldane-Anderson model. } \label{fig4}
\end{figure}

Figures 4(a) and (b) show $\langle M_1^z M_2^z\rangle$ versus
$k_0R$, where $R$ is the impurity separation, for the two-impurity
Anderson model for half-filled metallic ($\mu=-6.0$) and
semiconductor ($\mu=0.1$) cases. In the metallic case, $\langle
M_1^zM_2^z\rangle$ exhibits RKKY-type oscillations with both FM
and AFM correlations depending on the value of $k_0R$. On the
other hand, for $\mu=0.1$, we observe FM correlations of which
range increases with $\beta$.

\begin{figure}[t]
\centering
\includegraphics[width=7cm,bbllx=55,bblly=75,bburx=365,bbury=310,clip]{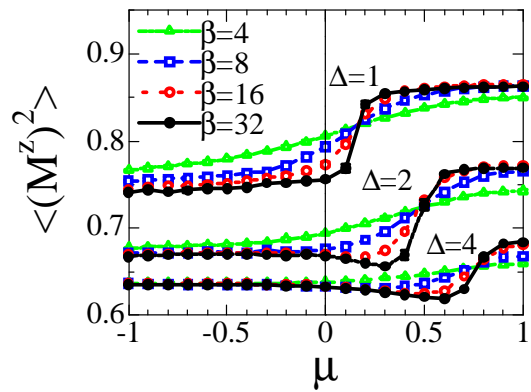}
\caption{ (Color online) Impurity magnetic-moment square $\langle
(M_z)^2\rangle$ vs $\mu$ at various $\beta$ for $\Delta=1$, 2 and
4 for the single-impurity Haldane-Anderson model. } \label{fig5}
\end{figure}

\begin{figure}[t]
\centering
\includegraphics[width=7cm,bbllx=55,bblly=80,bburx=365,bbury=320,clip]{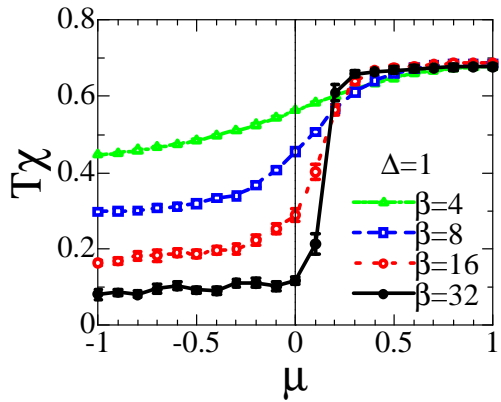}
\caption{ (Color online) $T\chi$ vs $\mu$ at various $\beta$ for
$\Delta=1$ for the single-impurity Haldane-Anderson model. }
\label{fig6}
\end{figure}

The remainder of the data shown for the 2D case in this section
are for the single-impurity Anderson model. In Fig. 5, we show
results on $\langle (M^z)^2 \rangle$ vs $\mu$ for various values
of $\Delta$. As $T$ is lowered, a step discontinuity develops in
$\langle (M^z)^2 \rangle$ near the gap edge. The location of the
discontinuity coincides with the location of IBS deduced from data
on $\langle M_1^z M_2^z \rangle$, the impurity single-particle
spectral weight $A(\omega)$, and the inter-impurity susceptibility
$\chi_{12}$ discussed previously \cite{Bulut}. Hence, Fig. 5 shows
that the local moment increases rapidly as the IBS becomes
occupied. Here, we also observe that the moment size decreases
with increasing hybridization, as expected. Figure 6 shows $T\chi$
vs $\mu$ for $\Delta=1$, where we observe that a step
discontinuity develops at the same location as in $\langle (M^z)^2
\rangle$ shown in Fig. 5(a). For $\mu\ltsim \omega_{IBS}$, $T\chi$
decreases with decreasing $T$ due to the screening of the impurity
moment by the valence electrons. However, for $\mu \gtsim
\omega_{IBS}$, the impurity susceptibility exhibits free-moment
behavior in agreement with the role of the IBS discussed above. We
note that determining the location of the IBS from $A(\omega)$ is
costly in terms of computation time. For this reason, in the
single-impurity case, it is more convenient to determine
$\omega_{IBS}$ from data on $\langle (M^z)^2\rangle$ versus $\mu$.
In the remaining sections, we will use this approach for
determining $\omega_{IBS}$.

Within the HF approximation and in 2D, the range of the FM
correlations is given by $( 16\pi \rho_0 \omega_{IBS} )^{-1/2}$
for $0< \mu< \omega_{IBS}$. This implies that the range decreases
with increasing $\omega_{IBS}$ and, hence, $\Delta$. We find that
the QMC results on the maximum range of the FM correlations are in
quantitative agreement with the values from the HF calculations
\cite{Bulut}. However, we also find that the HF approximation
underestimates the value of $\omega_{IBS}$. For example, for
$\Delta=1$, HF yields $\omega_{IBS}=0.017$, which is about an
order of magnitude smaller. Hence, within HF, the long-range FM
correlations are restricted to a narrow range of $\mu$.

\subsection{Impurity-host correlations}

\begin{figure}[t]
\centering
\includegraphics[width=5.7cm,bbllx=55,bblly=80,bburx=365,bbury=320,clip]{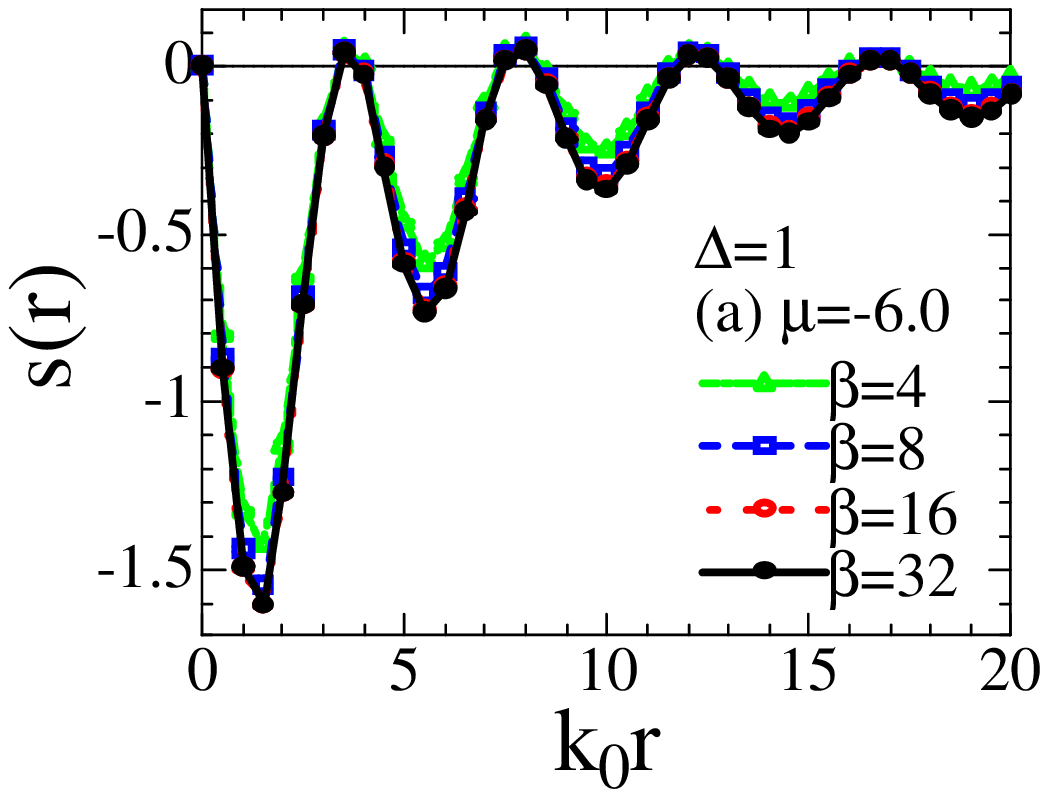}
\includegraphics[width=5.7cm,bbllx=55,bblly=80,bburx=365,bbury=320,clip]{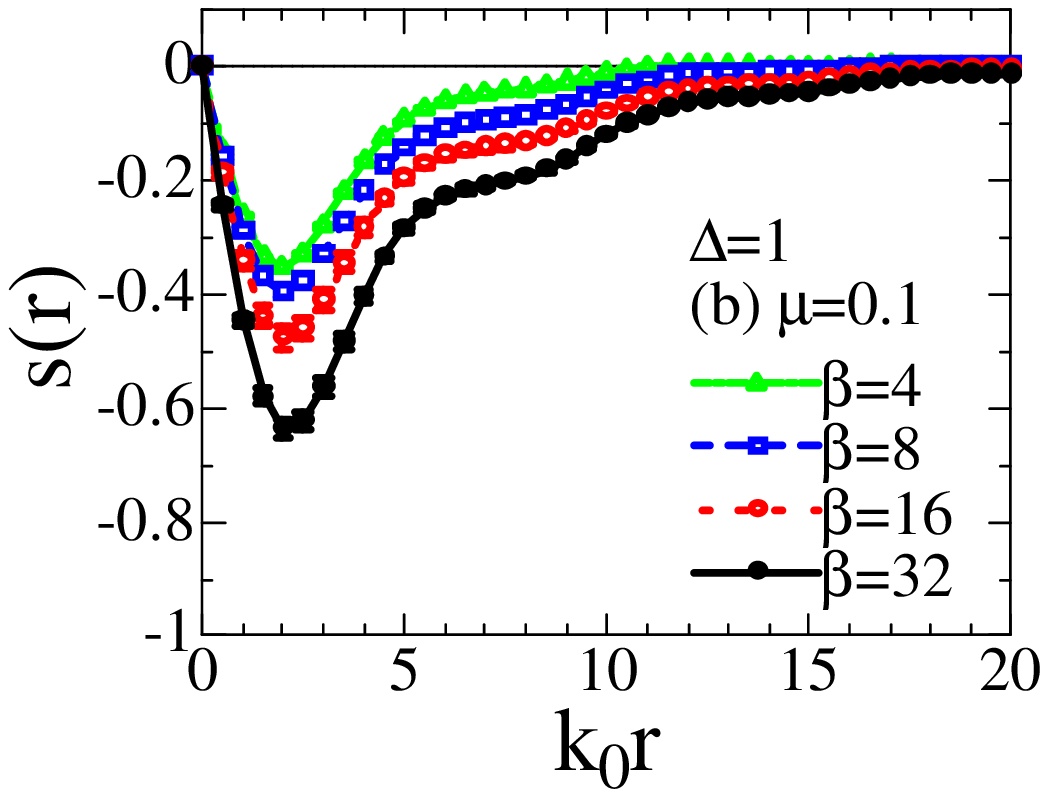}
\includegraphics[width=5.7cm,bbllx=55,bblly=80,bburx=365,bbury=320,clip]{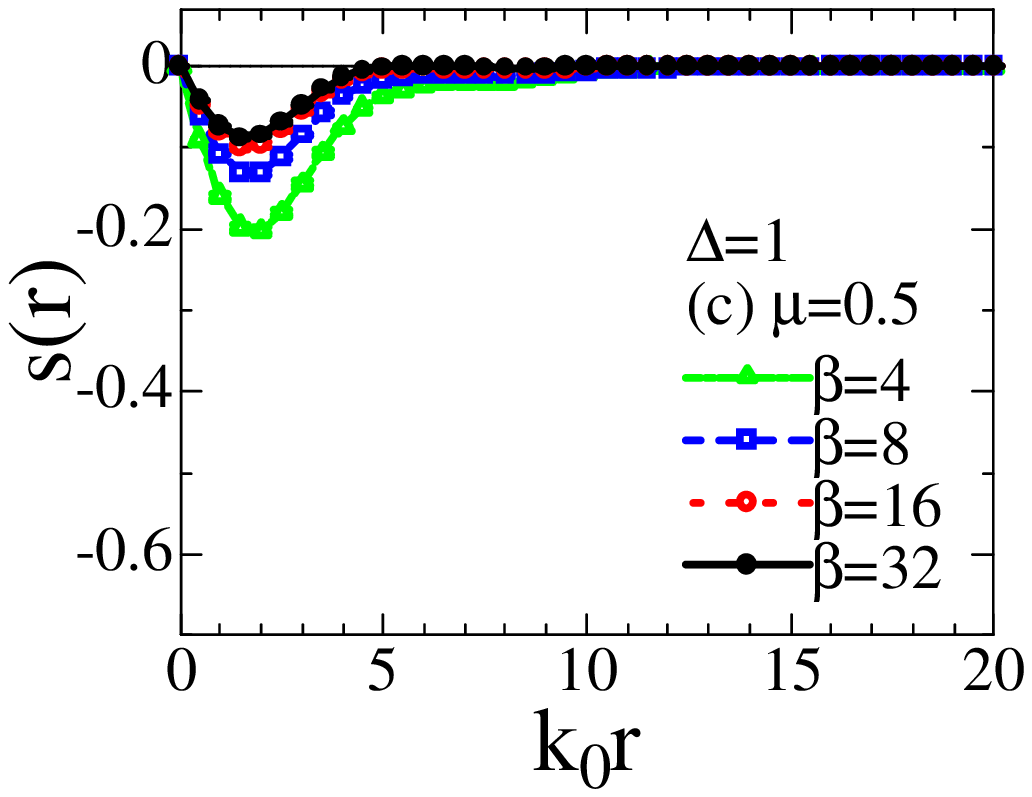}
\caption{ (Color online) Impurity-host magnetic correlation
function $s(r)$ vs $k_0r$ at various $\beta$ for (a) $\mu=-6.0$,
(b) $\mu=0.1$, and (c) $\mu=0.5$ for the single-impurity
Haldane-Anderson model. } \label{fig7}
\end{figure}

In this section, we discuss the magnetic correlations between the
impurity and the host. In addition, we show results on the induced
charge oscillations around the impurity site.

Figures 7(a)-(c) show the impurity-host magnetic correlation
function $s(r)$ defined by
\begin{equation}
s(r)= \frac{2\pi k_0r}{n_0} \langle M^z m^z(r)\rangle
\end{equation}
versus $k_0r$ for the single-impurity Anderson model for
$\mu=-6.0$ (half-filled metallic), $\mu=0.1$ (semiconductor with
IBS unoccupied), and $\mu=0.5$ (semiconductor with IBS occupied).
Here, $n_0$ is the electron density and $r$ is the distance from
the impurity site. For $\mu=0.1$, the coupling between the
impurity and host spins is AFM for all values of $k_0r$, while,
for $\mu=-6.0$, $s(r)$ exhibits RKKY-type $2k_F$ oscillations. We
also note that, for the metallic case, the magnetic correlations
saturate before reaching $\beta=32$. Comparison of Fig. 7(b) and
Fig. 4(b) for $\mu=0.1$ show that the AFM impurity-host coupling
produces the FM correlations between the impurities.

\begin{figure}
\centering
\includegraphics[width=7cm,bbllx=55,bblly=80,bburx=365,bbury=320,clip]{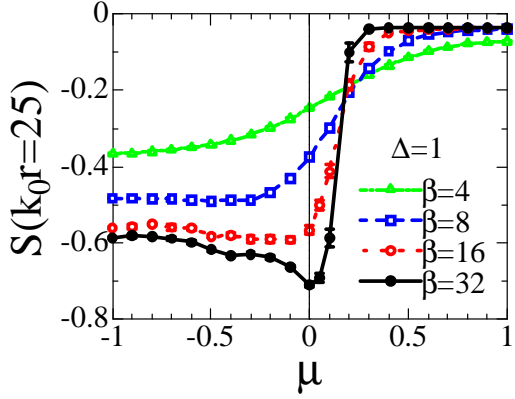}
\caption{ (Color online) $S(k_0 r=25)$ vs $\mu$ at various $\beta$
for the single-impurity Haldane-Anderson model. } \label{fig8}
\end{figure}

As seen in Fig. 7(c), when $\mu$ is increased to 0.5, the AFM
correlations between the impurity and host spins become much
weaker. This is because the IBS becomes occupied for $\mu > 0.1$.
Within the HF approximation \cite{Ichimura,Takahashi}, the FM
interaction between the impurities is mediated by the
impurity-induced polarization of the valence electron spins, which
exhibit an AFM coupling to the impurity moments. The impurity-host
hybridization also induces host split-off states at the same
energy as the IBS. When the split-off state becomes occupied, the
spin polarizations of the valence band and the split-off state
cancel. This causes the long-range FM correlations between the
impurities to vanish. These QMC and HF results emphasize the role
of the IBS in determining the range of the magnetic correlations
for a semiconductor host.

The total magnetic coupling of the impurity magnetic moment to the
host is obtained from
\begin{equation}
S(k_0r) = \int_0^{k_0r} d(k_0r') s(r')
\end{equation}
Figure 8 shows $S(k_0r=25)$ vs $\mu$ for $\Delta=1$ at various
$\beta$. Here, we see that the impurity becomes magnetically
decoupled from the host when $\mu \gtsim \omega_{IBS}$.

\begin{figure}
\centering
\includegraphics[width=5.7cm,bbllx=55,bblly=85,bburx=365,bbury=325,clip]{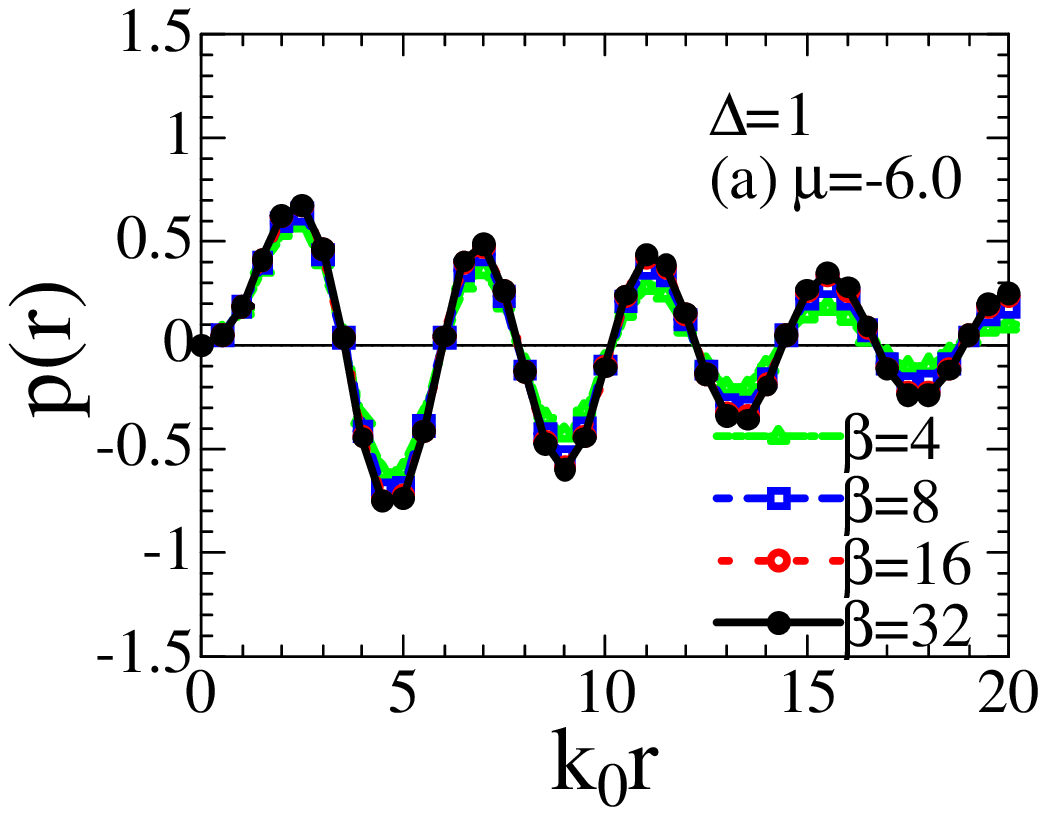}
\includegraphics[width=5.7cm,bbllx=55,bblly=85,bburx=365,bbury=325,clip]{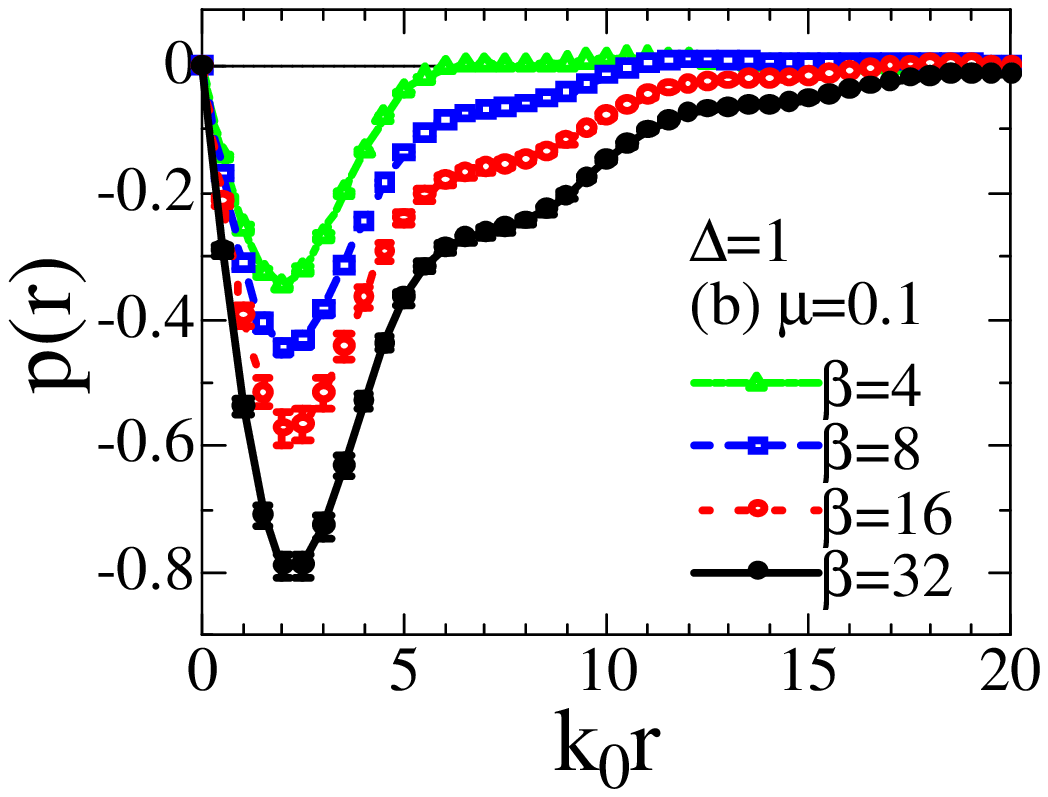}
\includegraphics[width=5.7cm,bbllx=55,bblly=85,bburx=365,bbury=325,clip]{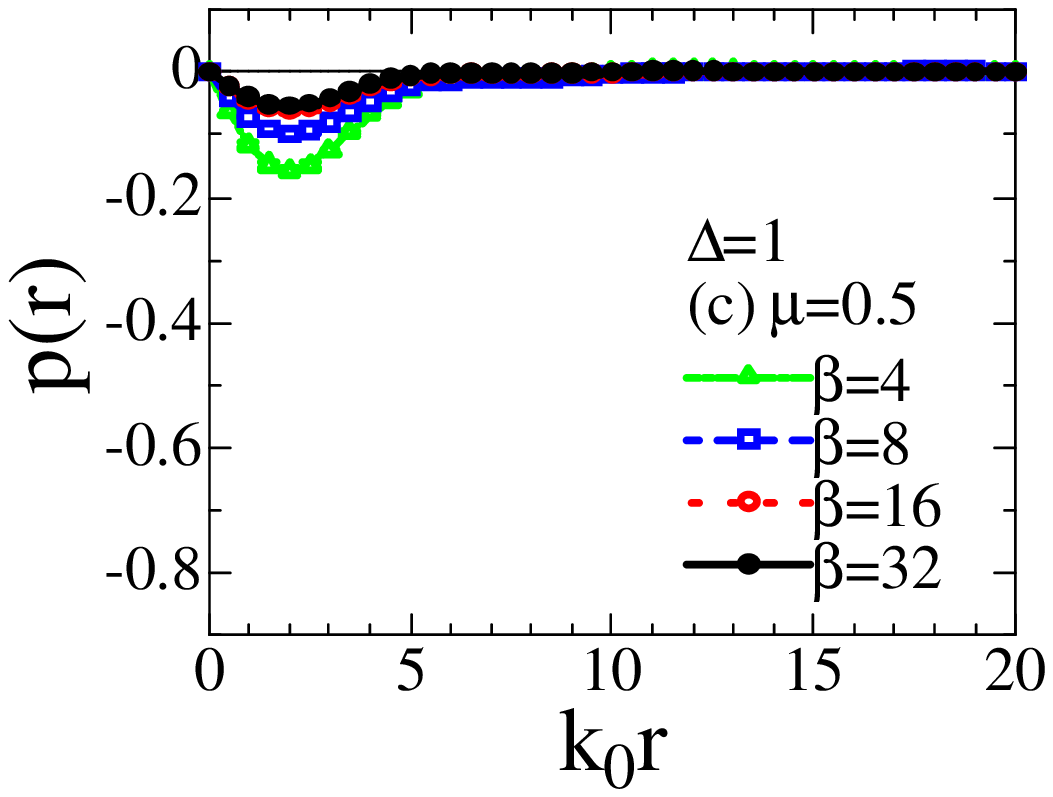}
\caption{ (Color online) $p(r)$ vs $k_0r$ for $\Delta=1$ and (a)
$\mu=-6.0$, (b) 0.1 and (c) 0.5 for the single-impurity
Haldane-Anderson model. } \label{fig9}
\end{figure}

Next, in Fig. 9, we show the modulation of the charge density
around the impurity. Here, we plot $p(r)$ vs $k_0r$, where $p(r)$
is defined by
\begin{equation}
p(r)=\sum_{\alpha=v,c} { 2\pi k_0r \over n_0} ( n_{\alpha}(r) -
n_{\alpha}(\infty))
\end{equation}
with $n_{\alpha}(r)=\sum_{\sigma} \langle
c^{\dagger}_{\alpha\sigma}(r) c_{\alpha\sigma}(r)\rangle$. For the
metallic case of $\mu=-6.0$, we observe long-range RKKY-type of
oscillations in $p(r)$. When $\mu=0.1$, the charge density around
the impurity is depleted up to $k_0R\approx 20$ at these
temperatures. This depletion represents the extended valence hole
which forms around the impurity. However, for $\mu=0.5$, the
induced charge density decreases significantly as $T$ is lowered,
because now the IBS is occupied. We next integrate $p(r)$,
\begin{equation}
P(k_0r) = \int_0^{k_0r} d(k_0r') p(r'),
\end{equation}
and plot $P(k_0r=25)$ vs $\mu$ in Fig. 10. We observe that the
total charge density around the impurity is most depleted when $0
\ltsim \mu \ltsim \omega_{IBS}$, which is due to the valence hole
induced around the impurity. In the metallic case, the induced
charge density is oscillatory and has a long range as we have seen
in Fig. 9(a). As $\mu$ approaches the gap edge, the electron
density around the impurity is depleted. However, when $\mu \gtsim
\omega_{IBS}$, we see that this depletion is cancelled by the
extended charge density of the split-off state.

\begin{figure}
\centering
\includegraphics[width=7cm,bbllx=55,bblly=85,bburx=365,bbury=325,clip]{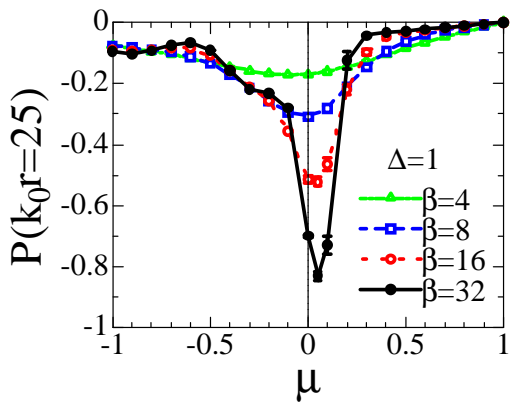}
\caption{ (Color online) $P(k_0 r=25)$ vs $\mu$ at various $\beta$
for the single-impurity Haldane-Anderson model. } \label{fig10}
\end{figure}

\section{QMC results for a 3D semiconductor host}

In this section, we discuss the three-dimensional case, where the
hybridization parameter $\pi V^2 N(0)$ vanishes at the gap edge
because of the vanishing of $N(0)$ of the pure host. Hence, the
impurity-host coupling near the gap edge is much weaker compared
to the 2D case. Consequently, the FM correlations between the
impurities have a shorter range. We find that the dimensionality
of the host material strongly influences the magnetic
correlations. In particular, we see that the IBS does not exist in
3D unless the hybridization is sufficiently large.

Here, we define hybridization as $\Delta=\pi \rho_0^* V^2$ where
$\rho_0^*$ is the density of states when the valence band is
half-filled, $\rho_0^*=k_0^3/(4\sqrt{2}\pi^2D)$. This choice
allows us to use comparable values for the hybridization matrix
element $V$ when we compare the 2D and 3D results. In the
following, we present results for $\Delta=1$, 2 and 4.

\begin{figure}[t]
\centering
\includegraphics[width=5.7cm,bbllx=55,bblly=85,bburx=360,bbury=325,clip]{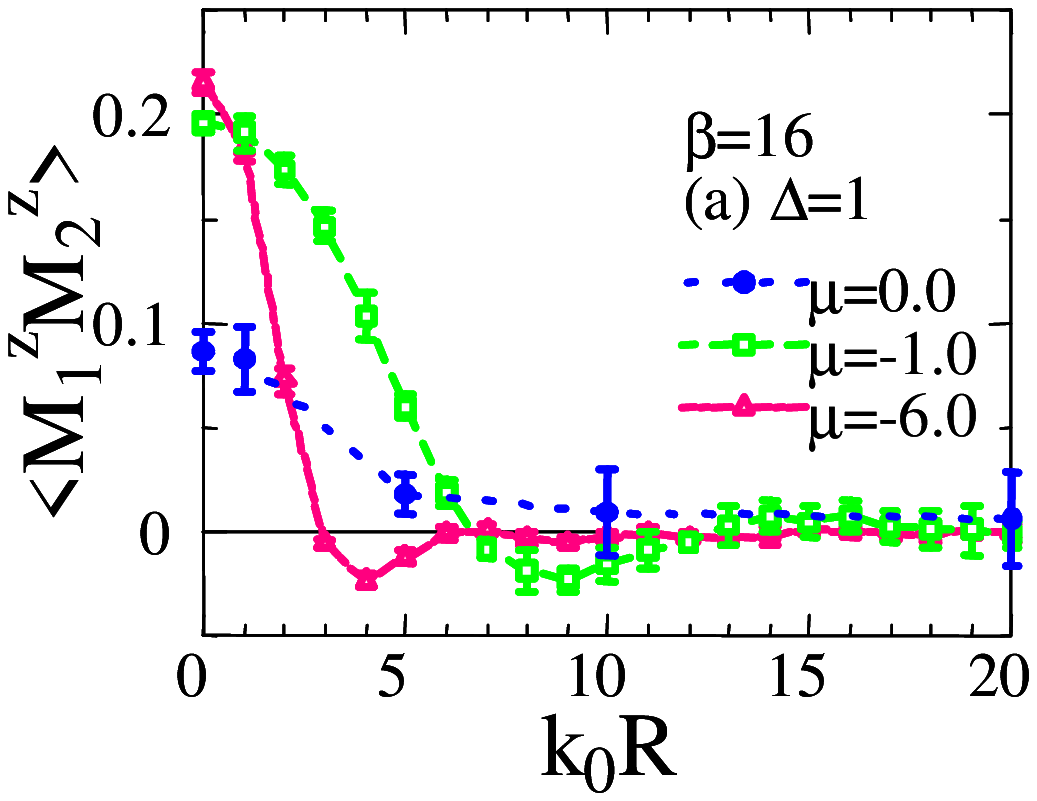}
\includegraphics[width=5.7cm,bbllx=55,bblly=85,bburx=360,bbury=325,clip]{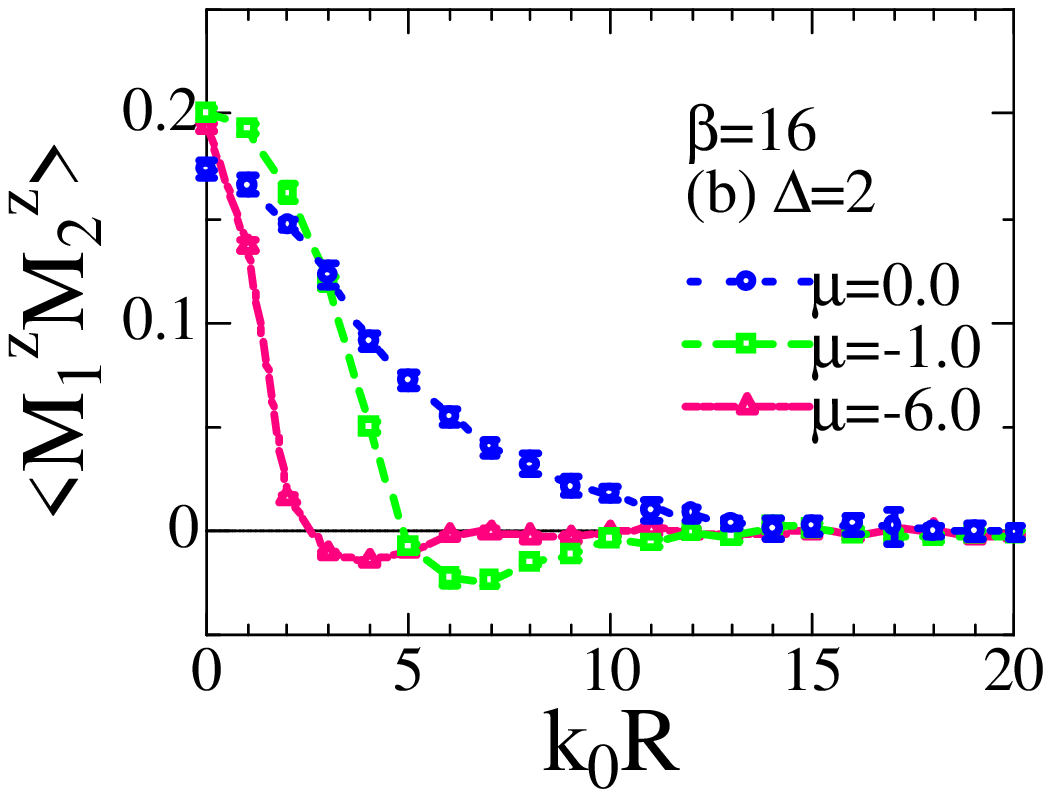}
\includegraphics[width=5.7cm,bbllx=55,bblly=85,bburx=360,bbury=325,clip]{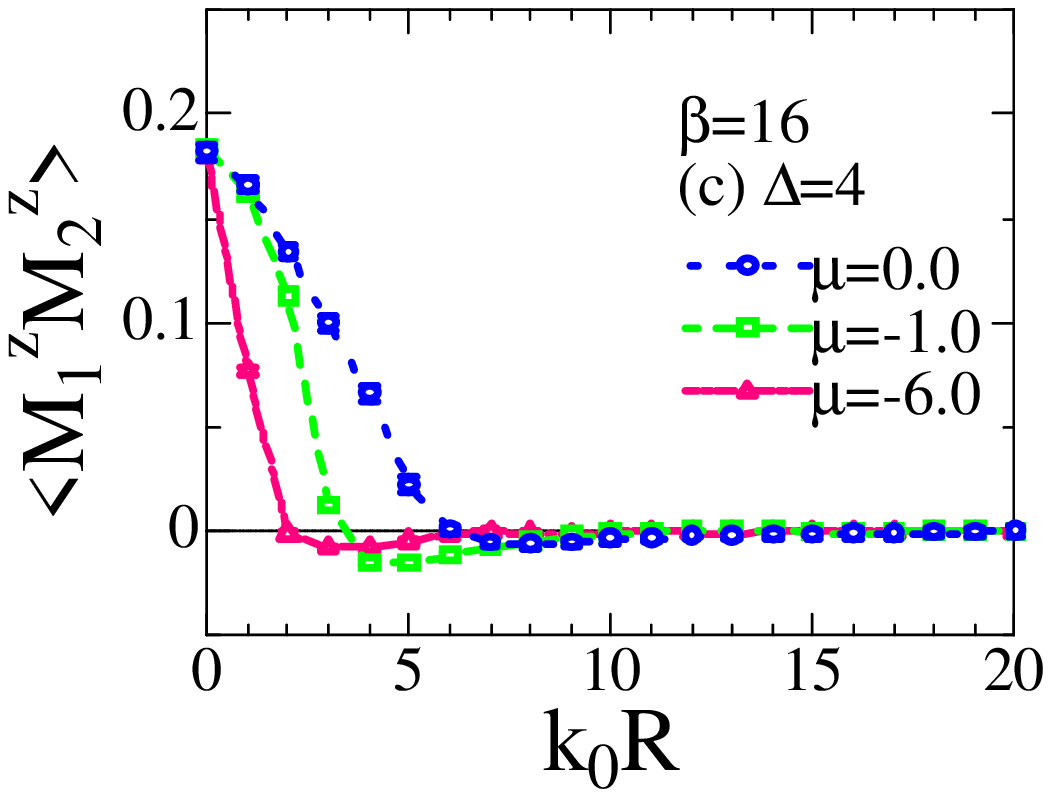}
\caption{ (Color online) $\langle M_1^z M_2^z \rangle$ vs $k_0 R$
at $\beta=16$ and various $\mu$  for hybridization (a)
$\Delta=1.0$, (b) 2.0 and (c) 4.0 for the two-impurity
Haldane-Anderson model in the 3D case. } \label{fig11}
\end{figure}

\begin{figure}[t]
\centering
\includegraphics[width=5.7cm,bbllx=55,bblly=80,bburx=360,bbury=315,clip]{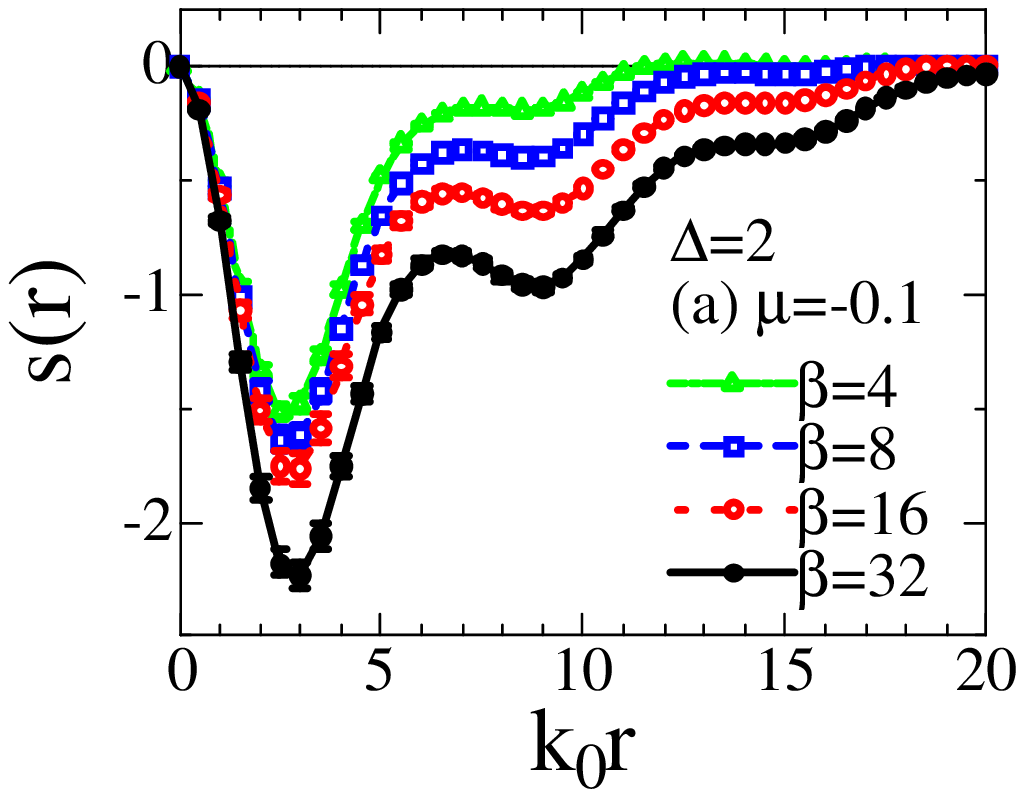}
\includegraphics[width=5.7cm,bbllx=55,bblly=80,bburx=360,bbury=315,clip]{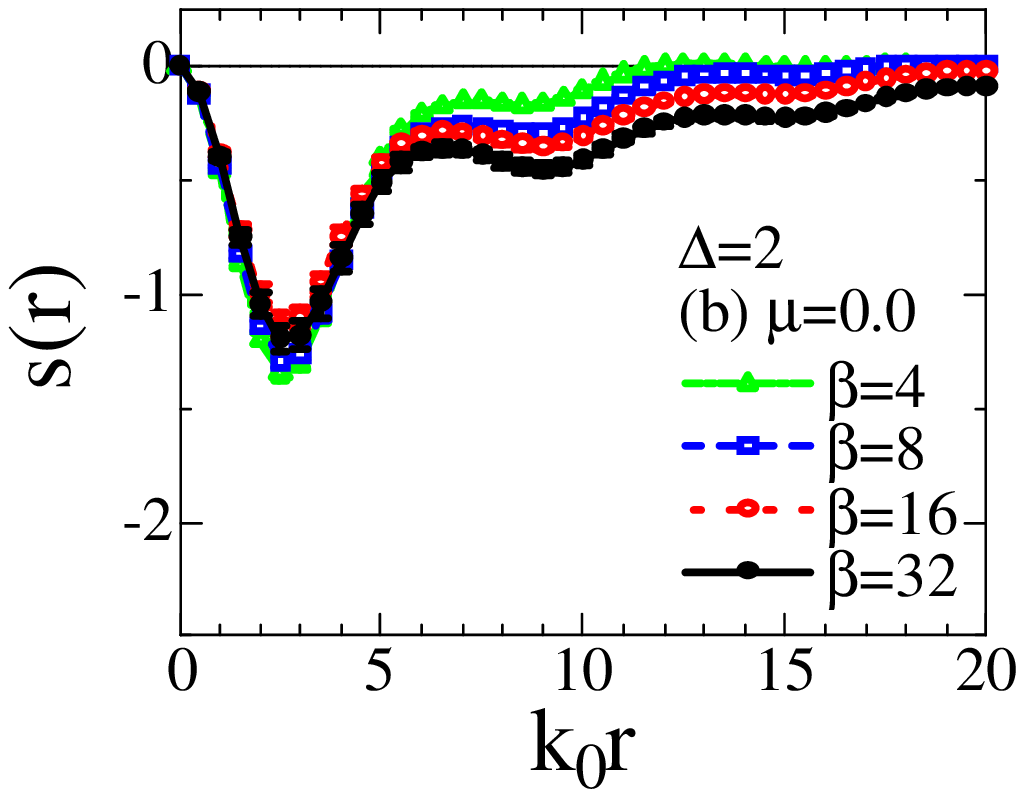}
\includegraphics[width=5.7cm,bbllx=55,bblly=80,bburx=360,bbury=315,clip]{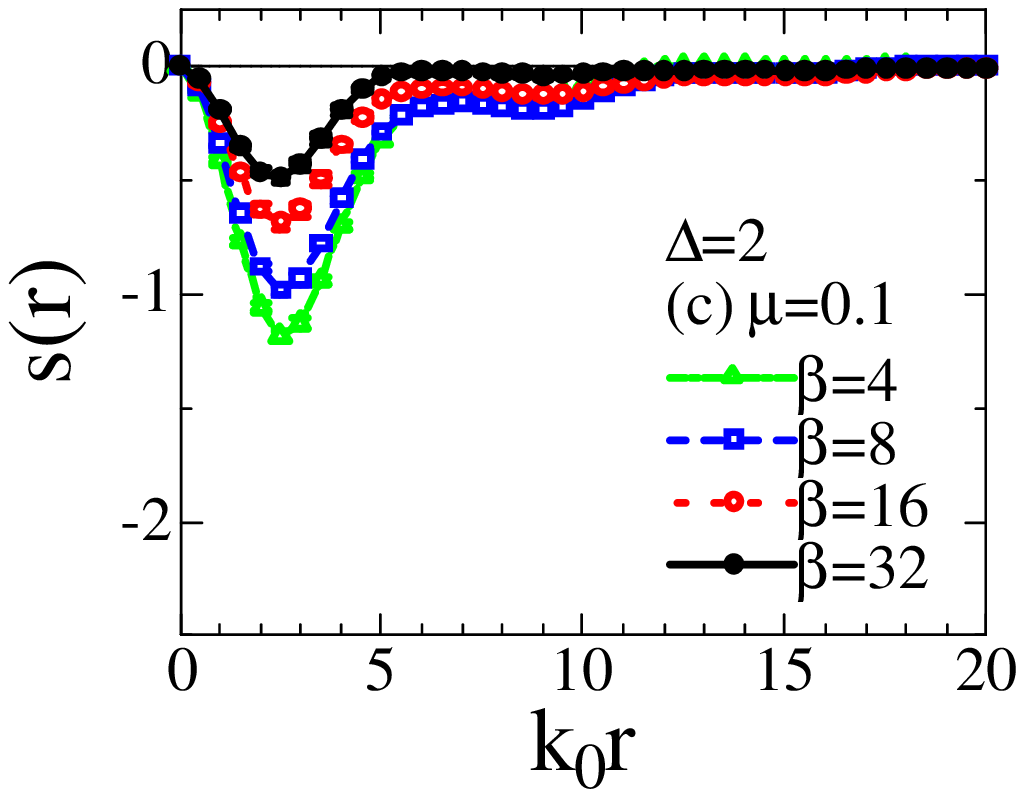}
\caption{ (Color online) $s(r)$ vs $k_0 r$ at various $\beta$ for
$\Delta=2$ and for (a) $\mu=-0.1$, (b) 0.0 and (c) 0.1 for the
single-impurity Haldane-Anderson model in a 3D host. }
\label{fig12}
\end{figure}

\begin{figure}[t]
\centering
\includegraphics[width=7cm,bbllx=55,bblly=75,bburx=360,bbury=310,clip]{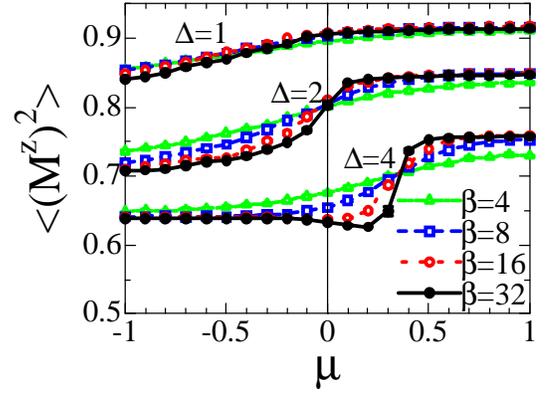}
\caption{ (Color online) Impurity magnetic-moment square $\langle
(M_z)^2\rangle$ vs $\mu$ at various $\beta$ for (a) $\Delta=1$, 2
and 4 for the single-impurity Haldane-Anderson model in a 3D host.
} \label{fig13}
\end{figure}

Figures 11(a)-(c) show $\langle M_1^z M_2^z \rangle$ vs $k_0R$ at
various values of $\mu$ for $\Delta=1$, 2 and 4. These results are
for $\mu=-6.0$, $\mu=-1.0$ and $\mu=0.0$. We observe that, for
$\Delta=1$, the FM correlations between the impurities weaken as
$\mu$ approaches the top of the valence band. On the other hand,
for $\Delta=2$ and 4, the FM correlations are stronger for
$\mu=0.0$. This is because, for $\Delta=1$, the IBS does not exist
in a 3D host, as we will see later in Fig. 13, which shows results
on $\langle (M^z)^2\rangle$ versus $\mu$.

Figure 12 shows the impurity-host magnetic correlation function
$s(r)$ vs $k_0r$ for the single-impurity case. In 3D, $s(r)$ is
defined by
\begin{equation}
s(r) = \frac{4\pi(k_0r)^2}{n_0} \langle M^z m^z(r)\rangle.
\end{equation}
We see that the impurity-host coupling weakens rapidly for $\mu
\gtsim 0.0$. These figures show that, in 3D and for $\Delta=2$,
the IBS is located at $\omega_{IBS}\approx 0.0$, which is
consistent with the results on $\langle (M^z)^2\rangle$ shown in
Fig. 13. In Fig. 13, for $\Delta=1$, we do not observe the
development of a discontinuity for temperatures down to
$\beta=32$. For $\Delta=2$, we observe the development of a step
centered at $\mu\approx 0.0$, as $\beta$ increases. For
$\Delta=4$, a step discontinuity at $\mu\approx 0.3$ is clearly
observed. So, in 3D the IBS exists only for sufficiently large
values of the hybridization matrix element $V$. This is consistent
with the dependence of the IBS on the dimensionality in the $U=0$
case. These results show that the dimensionality of the host
material strongly influences the magnetic properties.

The results presented in Sections 3 and 4 show that the density of
states of the pure host at the gap edge and the value of the
hybridization matrix element are crucial in determining presence
and the location of the IBS. The IBS is in turn important in
determining the magnetic properties of the systems when transition
metal impurities are substituted into a semiconductor host. This
means that the electronic state of the pure host material will
also be crucial in determining the magnetic properties. In the
next section, we explore the consequences of a more realistic band
structure for a GaAs host using the tight-binding approximation.

\section{QMC results for the tight-binding model of a
$\textbf{Mn}$ impurity in $\textbf{GaAs}$}

In this section, we present QMC data obtained within the
tight-binding model of a Mn impurity in GaAs, which was introduced
in Section. 2.B. Here, we have performed the QMC calculations
keeping all three of the Mn $t_{2g}$ orbitals in Eq.~(1), hence
this approach includes the multi-orbital effects except for the
Hund coupling. In addition, we use a more realistic description of
the semiconductor bands $\varepsilon_{{\bf k}\alpha}$ for GaAs
compared to that of Sections 3 and 4. Furthermore, the
hybridization $V_{{\bf k}\alpha,\xi}$ is determined by the
tight-binding approach using parameters consistent with
photoemission measurements on Mn in GaAs, instead of being a free
parameter. The following QMC results are obtained for the
Slater-Koster parameters $(pd\sigma)=-1.14$ eV, $(sd\sigma)=0$ eV,
and $(pd\pi)=(pd\sigma)/(-2.16)$.

Figure 14 shows $\langle (M^z)^2 \rangle$ and $T\chi$ versus $\mu$
for the $xy$ orbital near the top of the valence band for the
single-impurity Haldane-Anderson model. Here, the development of a
step discontinuity in the semiconductor gap is clearly seen as $T$
decreases down to 180 K. At low $T$ and for $(pd\sigma)=-1.14$ eV,
the inflection point of the discontinuity occurs at 100 meV, which
implies that $\omega_{IBS}=100$ meV. This is in good agreement
with the experimental value of 110 meV, especially if we note that
the estimate of $(pd\sigma)$ from the photoemission experiments is
$-1.1$ eV.

\begin{figure}[t]
(a)\includegraphics[width=5.5cm,bbllx=55,bblly=80,bburx=365,bbury=320,clip]{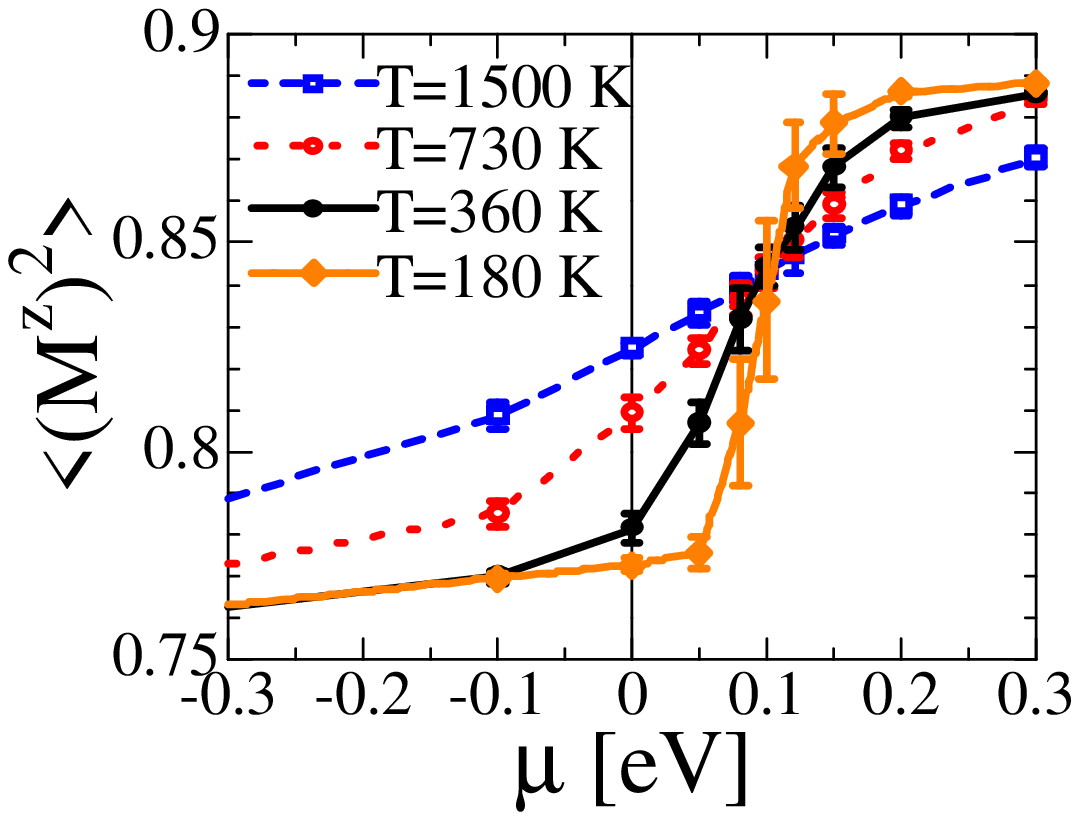}
(b)\includegraphics[width=5.5cm,bbllx=55,bblly=80,bburx=365,bbury=320,clip]{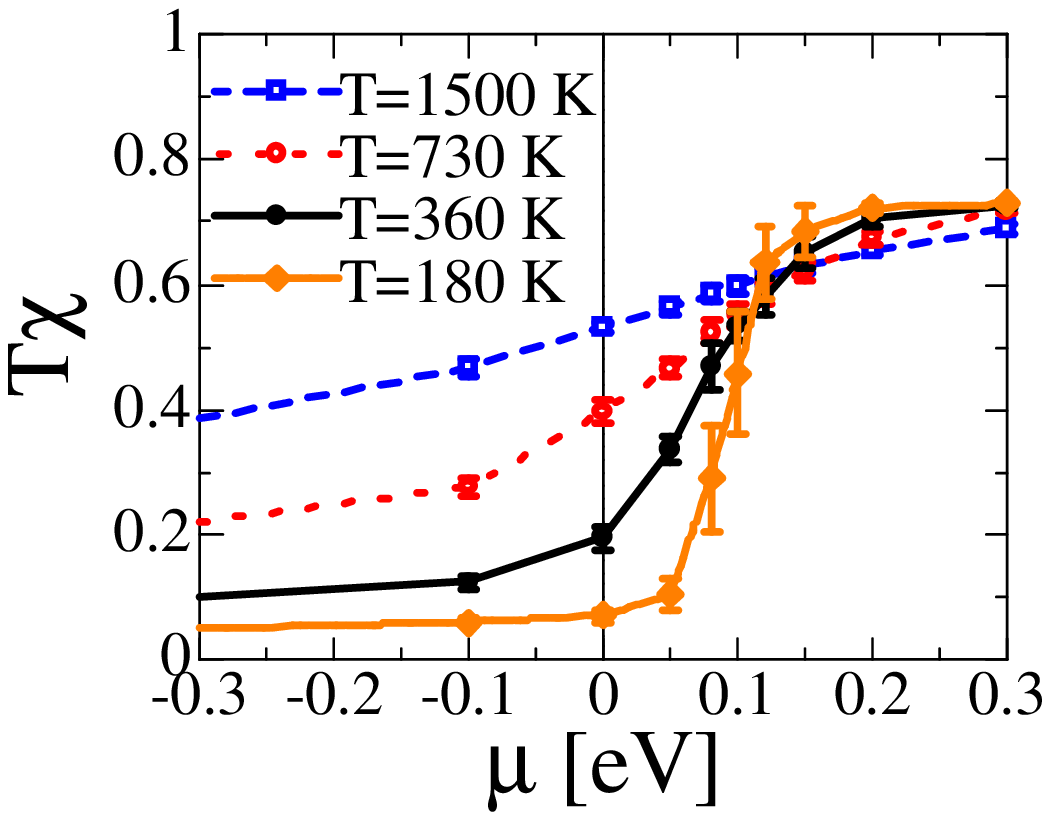}
\caption{ (Color online) (a) $\langle (M^z)^2\rangle$ and (b)
$T\chi$ versus $\mu$ for a Mn $xy$ orbital obtained using the
tight-binding model. Here, the top of the valence band is located
at $\mu=0$.} \label{fig14}
\end{figure}

In order to understand how sensitively $\omega_{IBS}$ depends on
the Slater-Koster parameters, we repeated these calculations for
different values of $(pd\sigma)$ keeping $(sd\sigma)=0$ and
$(pd\pi)=(pd\sigma)/(-2.16)$. For $(pd\sigma)=-1.4$ eV, we found
$\omega_{IBS}\approx 300$ meV, while for $(pd\sigma)=-1.0$ we
obtained a more smeared discontinuity in $\langle (M^z)^2 \rangle$
versus $\mu$ centered at $\mu\approx 0$.

\begin{figure}[t]
\centering
\includegraphics[width=5.7cm,bbllx=55,bblly=85,bburx=375,bbury=325,clip]{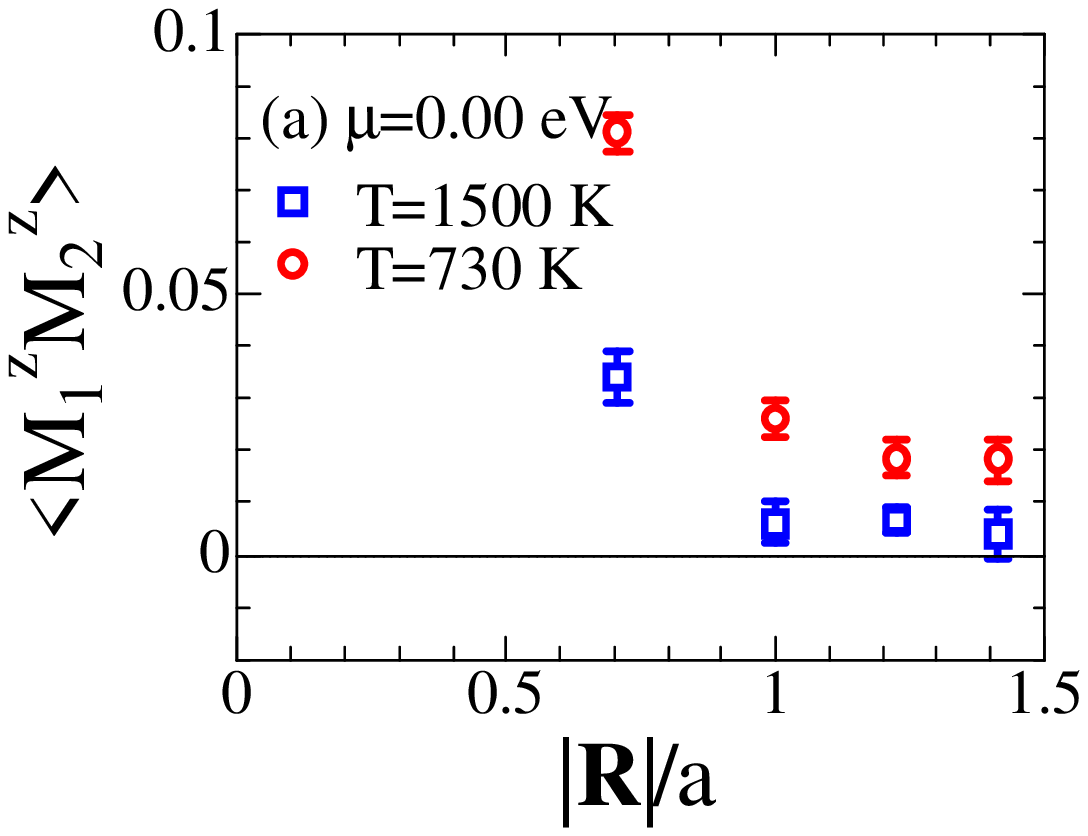}
\includegraphics[width=5.7cm,bbllx=55,bblly=85,bburx=375,bbury=325,clip]{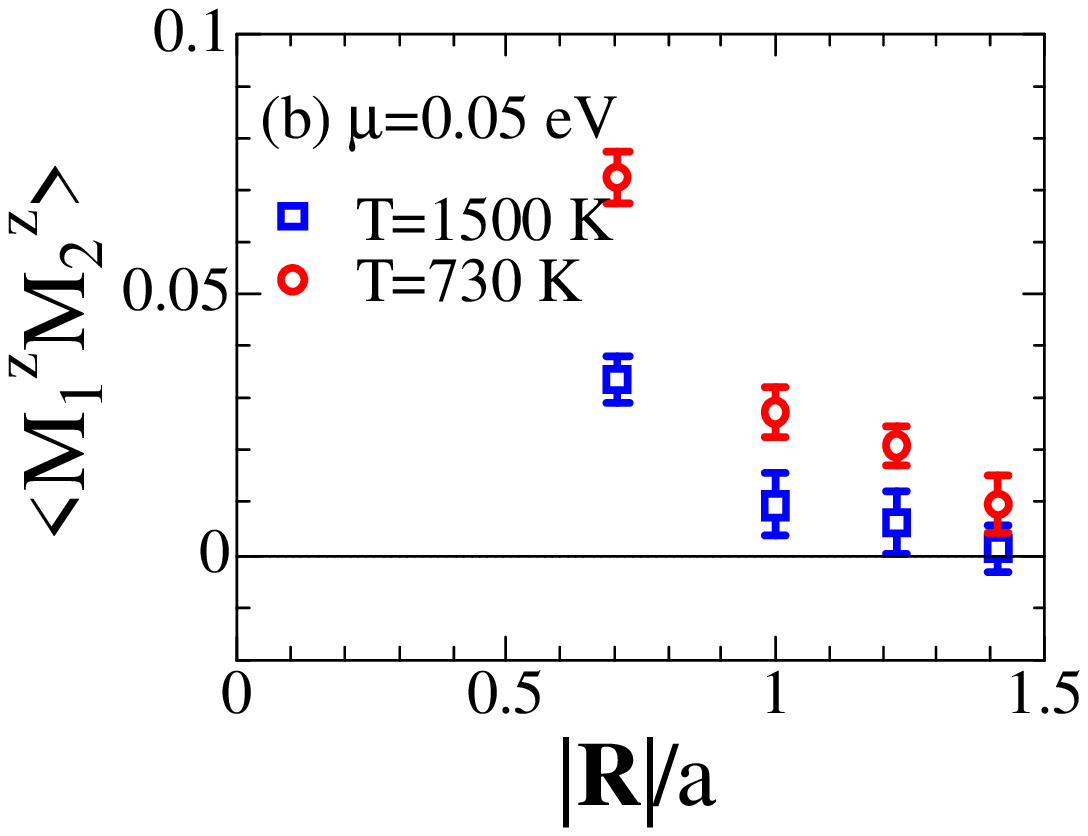}
\includegraphics[width=5.7cm,bbllx=55,bblly=85,bburx=375,bbury=325,clip]{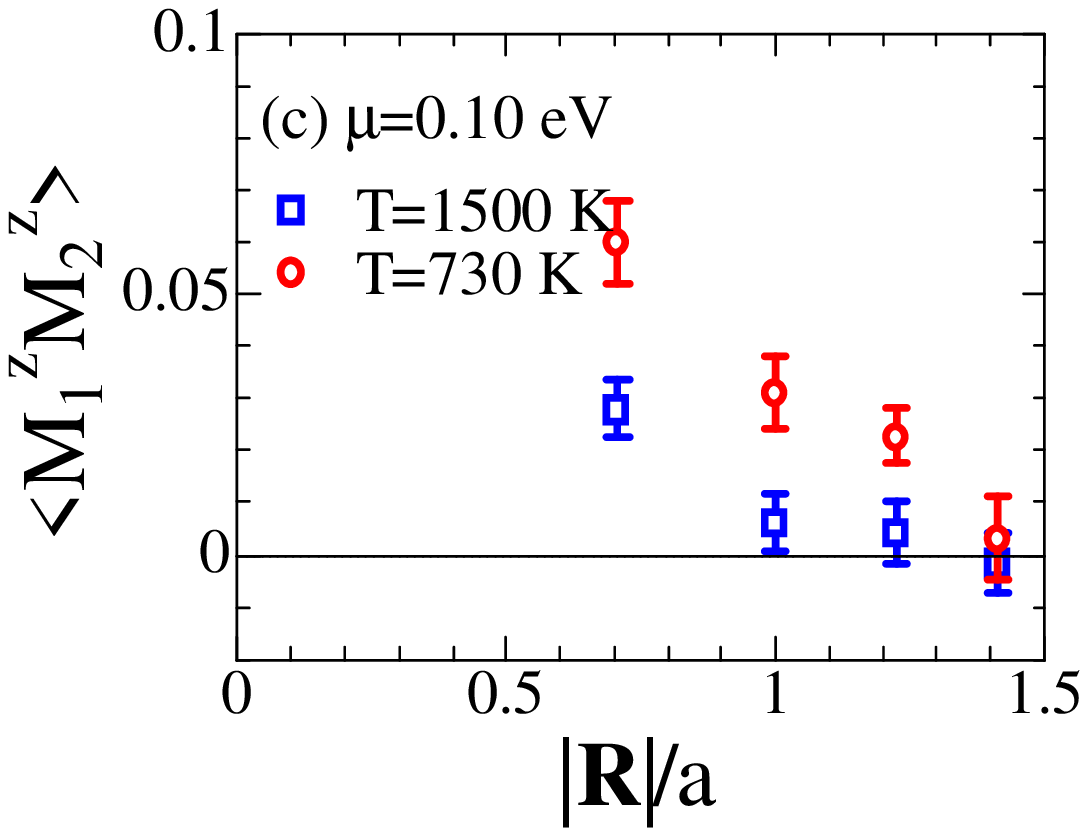}
\caption{ (Color online) Inter-impurity magnetic correlation
function $\langle M_1^z M_2^z \rangle$ vs $R/a$ at various $T$ for
(a) $\mu=0.0$ eV, (b) 0.05 eV and (c) 0.10 eV for the two-impurity
Haldane-Anderson model within the tight-binding model. Here,
$M^z_i$ is the magnetization operator at the Mn $xy$ orbital. }
\label{fig15}
\end{figure}

Next in Fig.~15, we display $\langle M_1^z M_2^z \rangle$ versus
the impurity separation in lattice units, $|{\bf R}|/a$, for the
$xy$ orbital in the case of two Mn impurities substituted into Ga
sites. Since we keep three orbitals at each Mn site, these
calculations are more costly in terms of computer time. In this
case, we present data for $T=1500$ K and 730 K. The Hirsch-Fye
algorithm for the single and two-impurity Anderson model does not
have the fermion sign problem. Hence, by using sufficient amount
of computer time it is possible, in principle, to extend the
calculation of $\langle M_1^z M_2^z \rangle$ to lower
temperatures. Even though the data on $\langle M_1^z M_2^z
\rangle$ are at high temperatures, the effect of the IBS is
observable. For $T=730$ K and $\mu=0.0$, the FM correlations
extend up to $1.5a$, however the FM correlations weaken as $\mu$
approaches $\omega_{IBS}$. We note that the spatial extent of the
Mn-Mn interaction in low-density (Ga,Mn)As was also considered by
using the tight-binding approximation and perturbative techniques
in Ref.~\cite{Tang}.

Finally, we discuss the correlations between the host electronic
spins and the magnetic moment of the $xy$ orbital at the Mn site
for the single-impurity case. In Fig.~16, we plot $\langle M^z
m^z({\bf r}) \rangle$ versus the impurity distance $|{\bf r}|/a$
for various values of $\mu$ and $T$. Here, we denote the
first-neighbor As site by As$^1$, the second-neighbor Ga site by
Ga$^2$, and so on. These results show that the development of the
AFM correlations between the magnetic moment of the $xy$ orbital
and the neighboring host electrons as $T$ is lowered down to 360
K. We also observe the weakening of  the AFM correlations as $\mu$
approaches $\omega_{IBS}$.

\begin{figure}[t]
\centering
\includegraphics[width=5.7cm,bbllx=55,bblly=75,bburx=295,bbury=245,clip]{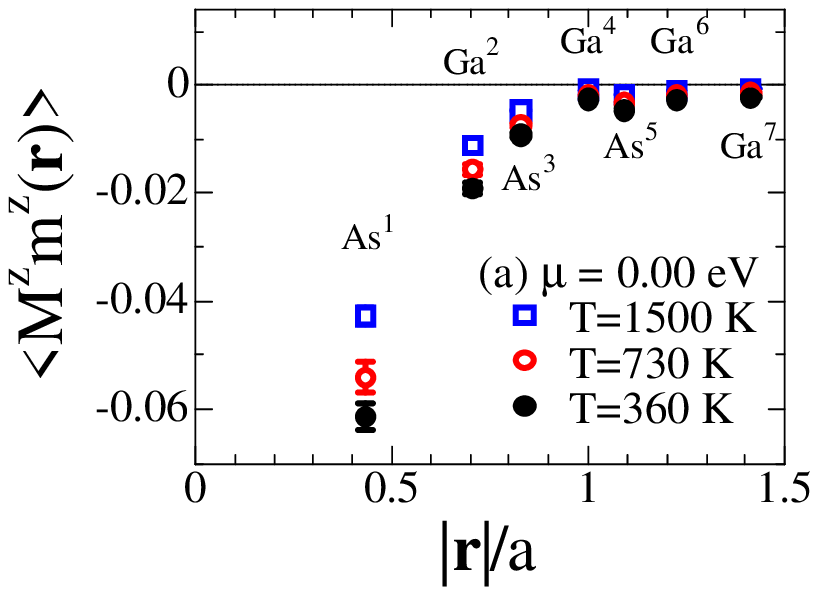}
\includegraphics[width=5.7cm,bbllx=55,bblly=75,bburx=295,bbury=245,clip]{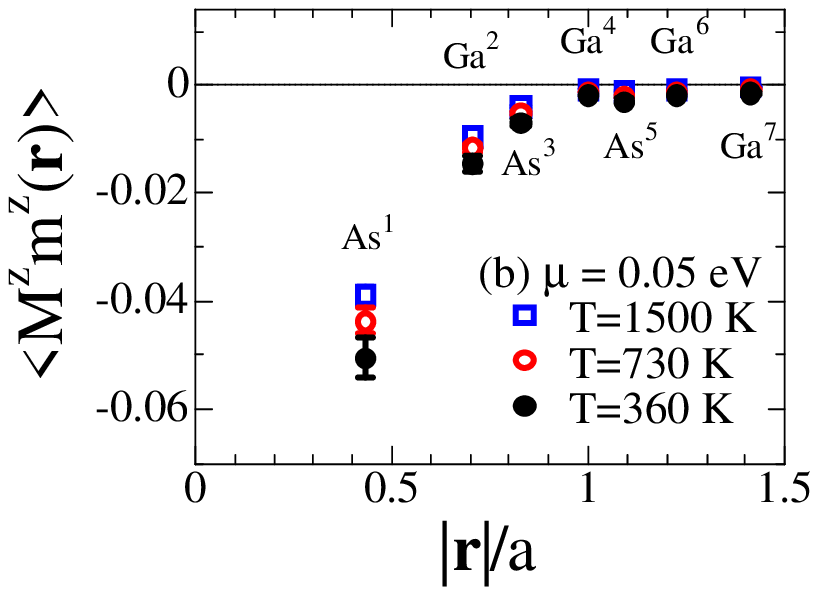}
\includegraphics[width=5.7cm,bbllx=55,bblly=75,bburx=295,bbury=245,clip]{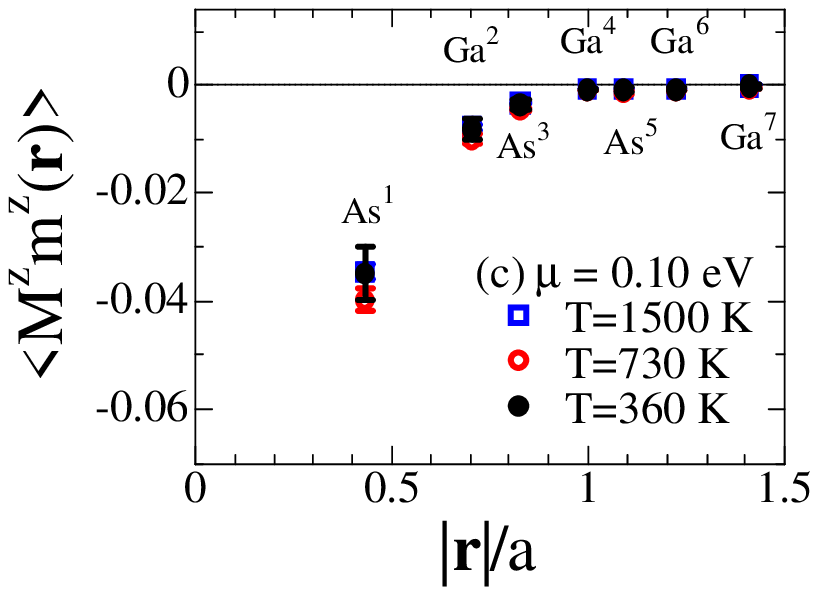}
\caption{ (Color online) Impurity-host magnetic correlation
function $\langle M^z m^z({\bf r}) \rangle$ vs $r/a$ at various
$T$ for (a) $\mu=0.0$ eV, (b) 0.05 eV and (c) 0.10 eV for the
single-impurity Haldane-Anderson model within the tight-binding
model. Here, $M^z$ is the magnetization operator at the Mn $xy$
orbital. } \label{fig16}
\end{figure}

Using the tight-binding results for $\varepsilon_{{\bf k}\alpha}$
and $V_{{\bf k}\alpha,\xi}$ as input, in this section, we have
performed the QMC simulations to study the magnetic properties
(Ga,Mn)As in the low-density limit. These results are similar to
those presented in Sections 3 and 4 for the single-orbital case
with simple band structure. In particular, we saw that, by using a
realistic model for the host band structure and host-impurity
hybridization of (Ga,Mn)As, it is possible to obtain an accurate
value for $\omega_{IBS}$. Such quantitative agreement supports the
physical picture described in this paper for the origin of the FM
correlations in DMS in the low-density limit.

\section{Discussion}

The computational approaches taken in this paper can be extended
to study the case of finite density of Mn impurities in GaAs.
Here, we have reported results for the single- and two-impurity
cases. However, the Haldane-Anderson model can also be studied for
a finite density of impurities using the Hirsch-Fye QMC algorithm.
This type of investigation could shed light on the metal-insulator
transition encountered in the DMS materials at finite density of
impurities. In addition, it could help to understand the nature of
the metallic state observed in (Ga,Mn)As for more than 2\% doping.

In the QMC results for the single-impurity case, we have observed
the extended nature of the induced charge density and the spin
polarization around the impurity site with a range $\ell_0$
determined by model parameters such as the hybridization, the bare
band structure of the host, the Coulomb repulsion, etc. In the
two-impurity case, when the polarization clouds overlap, the QMC
results showed that FM correlations develop between the
impurities. Then, the interesting question is what happens for
finite density of impurities as the average separation of the
impurities becomes comparable to $\ell_0$. Does the
Haldane-Anderson model capture the insulator-metal transition
observed in (Ga,Mn)As? In addition, is it possible to describe the
electronic properties of the metallic state bordering the
insulator-metal transition as that of a disordered valence band
with weak correlation effects? Or, does the metallic state retain
more properties of the single-impurity case so that the IBS and
the split-off states continue to play roles even though they
emerge with the valence band? In this case, what is the nature of
the magnetic correlations among the impurities? It would be
interesting to study this problem even in the simplest case where
each impurity has one orbital and the density of states of the
bare host and the hybridization are constants as discussed in
Section 3. By extending the QMC calculations to the finite-density
case, we think that it would be possible to address the question
of whether the metallic state bordering the insulator-metal
transition has unusual electronic properties.

\section{Summary}

In this paper, we studied the single- and two-impurity
Haldane-Anderson models using the Hirsch-Fye QMC technique. Our
purpose was to develop a microscopic understanding of the
low-doping insulating phase of the DMS material (Ga,Mn)As. We
first discussed the simple case where each impurity consists of
one localized orbital and the host band structure is described by
two and three-dimensional quadratic bands. We have presented QMC
results on the inter-impurity and impurity-host magnetic
correlations and the charge density around the impurity site. We
observed that the presence and the occupation of the IBS is
important in determining the magnetic correlations. We saw that
long-range FM correlations, which are induced by the AFM coupling
of the valence electrons to the impurity moment, develop between
the impurities when the chemical potential is between the valence
band and the IBS. We have also showed that, in the low-doping
limit, the magnetic correlations between the impurities do not
exhibit RKKY-type oscillations in a semiconductor. These
calculations also displayed how the model parameters determine the
magnetic correlations.

In order to develop a more realistic model of the low-doping
insulating phase of (Ga,Mn)As, we used the tight-binding
approximation to map the system to the Haldane-Anderson
Hamiltonian. In this approach, we used tight-binding parameters
determined from the photoemission experiments and kept all three
of the Mn $t_{2g}$ orbitals in the QMC calculations. The resulting
value for $\omega_{IBS}$ is close to the experimental value of 110
meV. We also observed that the magnetic correlations weaken as the
IBS becomes occupied. These results are useful for developing a
microscopic understanding of the low-doping insulating phase of
(Ga,Mn)As.

\begin{acknowledgements}
This work was supported by the NAREGI Nanoscience Project and a
Grant-in Aid for Scientific Research from the Ministry of
Education, Culture, Sports, Science and Technology of Japan, and
NEDO.
\end{acknowledgements}

\end{document}